\documentclass{article}
\usepackage{graphicx} % Required for inserting images

\usepackage[T1]{fontenc}
\usepackage[utf8]{inputenc} % Encodage
\usepackage[french, english]{babel}
\usepackage[cyr]{aeguill}

\usepackage{upgreek} % Lettres grecques droites

\usepackage[colorlinks=wrong]{hyperref}

\usepackage{xspace}

\usepackage{lmodern} % Mode police vectorielle 

\usepackage[top=4.cm, bottom=4.cm, left=4.cm, right=3.cm,headheight=17pt,,includehead,includefoot,heightrounded]{geometry} % recalcule des marges
\usepackage{emptypage} % Pages blanches non numérotées

\usepackage{appendix}

\usepackage{subcaption} % Sous-figures (obsolète)
\usepackage{graphicx} % Insérer images
\usepackage{caption} % Légendes
\usepackage{wrapfig} % Insérer images (mieux vaut faire autrement)
\usepackage{here} % Gérer la position des images

\usepackage{breakcites} % Coupe les citations longues

\usepackage{float}
\usepackage{array}

\usepackage{tikz} % Complexes de chaine et tout
\usepackage{tikz-cd}
\usetikzlibrary{arrows,calc,chains,matrix,positioning,scopes} % Complexes de chaine

\usepackage{tabularx}
\usepackage{multirow}
\usepackage{multicol}
\usepackage{bigdelim}

\usepackage{bbm} % Chiffres avec double barre

\usepackage{listings} % Citer du code

\usepackage{latexsym}
\usepackage{amsthm} % Mise en forme théorèmes
\usepackage{amsmath} % Formules mathématiques
\usepackage{amsfonts}
\usepackage{amsxtra}
\usepackage{yhmath}
\usepackage{amssymb} % Symboles
\usepackage{mathrsfs} % Objets mathématiques supplémentaires 
\usepackage{esint} % Symboles d'intégrales multiples
\usepackage{empheq}
\usepackage{stmaryrd}
\usepackage{blkarray} % annotations lignes et colonnes matrices

\usepackage{cleveref}% References croisées 
\crefformat{footnote}{#2\footnotemark[#1]#3}

\usepackage{cancel} % Barrer des expressions mathématiques
\usepackage{keyval}

\usepackage{comment}
\newcommand{\abs}[1]{\left\lvert#1\right\rvert}

\newcommand{\determinant}[2][]{\mathrm{det}_{#1}\mathopen{}\left(#2\right)}

\newcommand{\slfrac}[2]{\left.#1\middle/#2\right.}

\DeclareSymbolFont{UPM}{U}{eur}{m}{n}
\DeclareMathSymbol{\uppartial}{0}{UPM}{"40}

\newcommand{\ensemblenombre}[1]{\mathbb{#1}}

\newcommand{\Z}{\ensemblenombre{Z}}
\newcommand{\Q}{\ensemblenombre{Q}}
\newcommand{\R}{\ensemblenombre{R}}

\theoremstyle{definition}

\newtheorem{exple}{Example}

\newcommand{\bgamma}{\boldsymbol{\gamma}}
\newcommand{\bh}{\boldsymbol{\eta}}

\newcommand{\bfA}{\mathbf{A}}
\newcommand{\bfL}{\mathbf{L}}
\newcommand{\bfK}{\mathbf{K}}
\newcommand{\bht}{\bh_{\boldsymbol{t}}}
\newcommand{\bfQ}{\mathbf{Q}}
\newcommand{\bgt}{\bgamma_{\boldsymbol{t}}}

\newcommand{\tikzmark}[1]{\tikz[baseline,remember picture]
\coordinate (#1) {};}

\usepackage[
backend=biber,
style=numeric-comp,
 maxcitenames=50,
 maxnames=10,
sorting=none
]{biblatex}

\AtEveryBibitem{%
\clearfield{url}%
    \clearfield{urlyear}%
\ifentrytype{book}{
    \clearfield{url}%
    \clearfield{urlyear}%
}{}
\ifentrytype{collection}{
    \clearfield{url}%
    \clearfield{urlyear}%
}{}
\ifentrytype{incollection}{
    \clearfield{url}%
    \clearfield{urlyear}%
}{}
}

\makeatletter
\def\ps@pprintTitle{%
  \let\@oddhead\@empty
  \let\@evenhead\@empty
  \let\@oddfoot\@empty
  \let\@evenfoot\@oddfoot
}
\makeatother

\title{Observables in $\mathrm{U}(1)^n$ Chern-Simons theory}

\begin{document}
        \pagestyle{empty}
	\setcounter{page}{0}
        \hspace{-1cm}
	\begin{center}
		{\Large {\bf Observables in $\mathrm{U}(1)^n$ Chern-Simons theory}}%
		\\[1.5cm]
		
		{\large Michail Tagaris and Frank Thuillier}
		\vspace{3mm}
		
		{\it LAPTh, Univ. Savoie Mont Blanc, CNRS, F-74000 Annecy, France, michail.tagaris@lapth.cnrs.fr and frank.thuillier@lapth.cnrs.fr}
	\end{center}
\vspace{3cm}

\centerline{{\bf Abstract}}
In this article, we will compute the expectation value of observables (which appear as Wilson loops) in  $\mathrm{U}(1)^n$ Chern-Simons theory for closed oriented $3$-manifolds. We will show how the various topological sectors of the observable affect the expectation value and confirm that it is a topological invariant. We will also exhibit in this case as well a form of the CS duality introduced in previous works. Finally, to complete the treatment of this theory, we will compute its zero modes and the equations of motion.
\vspace{2cm}

\vfill
	\newpage
	\pagestyle{plain} 

\section{Introduction}
In recent articles, we explored the generalization of the $\mathrm{U}(1)$ Chern-Simons (CS) partition function, to the $\mathrm{U}(1)^n$ case\cite{kim_u1n_2025} which introduces some new concepts such as the inclusion of $\mathrm{U}(1)^n$ BF theory into a $\mathrm{U}(1)^{2n}$ CS and the CS duality, which shows a connection between two Chern-Simons partition functions with different parameters. We also explored the $\mathrm{U}(1)^n$ Reshetikhin-Turaev construction \cite{tagaris_reshetikhin-turaev_2025} and showed, in this case as well, its connection with Chern-Simons. 

In this article, we wish to explore the behavior of the observables on such a CS theory and analyze the above concepts with regard to observables as well. 
In the next section, we will perform the full computation of the expectation value while taking care of special cases that arise. At the end of the section we will also discuss how CS duality manifests in the observables as well. Then we will present a couple of examples that showcase how the computation and the CS duality work in actuality. Finally, in appendix \ref{Appendix} we also compute some other quantities related to the CS action, namely the zero modes and the equations of motion. Before we begin, there is an important note about the computation of the expectation value. We will be making use of Dehn surgery to represent the manifold and the observables in $S^3$ to make our computations easier. That being said, Dehn surgery is not necessary for any of these calculations.

\section{Expectation value of $\mathrm{U}(1)^n$ observables}
We will begin with a short reminder about the $\mathrm{U}(1)$ CS partition function. We start with a closed oriented $3$-manifold $M$. Since $\mathrm{U}(1)$ is topologically not simply connected, the fibre bundle corresponding to such a gauge group is not generally trivializable. This means that, in general, our fields are not globally defined over the whole manifold. To fix this issue, we have to consider only local fields and make use of the Deligne-Beilinson cohomology \cite{deligne_theorie_1971,beilinson_higher_1985,brylinski_loop_1993,esnault_deligne-beilinson_1988}. We define the $\mathrm{U}(1)$ CS action by \[ 
	S[{A}] = \oint_M {A} \star {A}. \]
    This is a functional that takes values in $\R / \Z$ and ${A}$ are elements of the first Deligne-Beilinson (DB) cohomology of $M$. In this case, $\star$ denotes the DB product. 

    As a reminder, representatives of DB classes consist of a $3$-tuple:
    \[
    A = [(X , \Lambda , \nu)],
    \] 
    where $X$, $\Lambda$ and $\nu$ are families of local fields. 
    The DB product between two such classes is commutative and for $B = [(Y, \Theta , \mu)]$ it is defined as:
    \[
    A \star B = [(X\wedge dY,\Lambda dY , \nu Y , \nu \Theta  , \nu\mu)]
    \]
    A comprehensive analysis of DB cohomology in $\mathrm{U}(1)$ Chern-Simons can also be found in \cite{calaque_deligne-beilinson_2015}. To generalize the theory to $\mathrm{U}(1)^n$we will consider an $n$-tuple of such fields $\mathbf{A}=({A}_1\dots {A}_n)$ and the action in this case will be:
    \[
    S_{\mathrm{CS}_{\mathbf{C}}}[{A}] = \oint_M \mathbf{A} ^\top \star_{C} \mathbf{A} := \sum_{i,j} C_{ij} \oint_M {A}_i \star {A}_j \, .
    \]
    $C_{ij}$ are integer numbers belonging to an integer matrix $\mathbf{C}$.
    To make sense of this integral, since it is only well defined modulo integers, we will have to evaluate it in a complex exponential. So the CS partition function looks like this:
    \[
\mathcal{Z}_{\mathrm{CS}_{\mathbf{C}}} = \frac{1}{\mathcal{N}} \int D \! \mathbf{A} \, e^{2 i \pi S_\mathbf{C}[\mathbf{A}]} \, ,
    \]
    with $\mathcal{N}$ a normalization defined such that it gets rid of an infinite sector that appears later in the computation. A small comment on the notation. We will use bold symbols like $\mathbf{A}$ when our quantities are defined in $\mathrm{U}(1)^n$ and thus are $n$-tuples of regular, non-bolded variables.

    \subsection{Computing the expectation value}
   If we have a Wilson loop represented by a closed path $\gamma$ inside the manifold, the observable associated with it in the $\mathrm{U}(1)$ case is the following:
    %We will now define the expectation value of observables. In the $\mathrm{U}(1)$, observables appear as Wilson loops:
    \[
     W(A,\gamma)=e^{2i\pi \int_\gamma A},
    \]
   The $\mathrm{U}(1)$ case through DB cohomology, has been studied extensively by \cite{guadagnini_path-integral_2014,mathieu_abelian_2018}. In the $\mathrm{U}(1)^n$ case this generalizes to
    \[
    W(\mathbf{A},\boldsymbol{\gamma})=e^{2i\pi \int_{\boldsymbol{\gamma}} \mathbf{A}},
    \]
    where now $\bgamma$ is an $n$-tuple of Wilson loops inside the manifold. By using the DB product, we can define the distributional DB class $\bh$ with support on $\bgamma$ such that  $\int_{\bgamma} \mathbf{A} = \int_M \bfA \star \bh$. The expectation value will then be
   % \[    \left< \left<  \right> \right>\]
    
\begin{align}
\label{eq:First_def_expectation_value}
\mbox{``} \langle \langle W_M (\mathbf{A}, \boldsymbol{\eta}) \rangle \rangle_{\mathrm{CS}_{\mathbf{C}}}
= \frac{1}{\mathcal{N}_{\mathrm{CS}_{\mathbf{C}}}(M)}
\int_{\left(H^1_{\mathrm{DB}}(M)\right)^{n}}\mathscr{D}\mathbf{A} 
e^{2\pi i \left( S_{\mathrm{CS}_{\mathbf{C}}}\left(\mathbf{A}\right)+\int_M \mathbf{A}\star \boldsymbol{\eta} \right)} \mbox{''}.
\end{align}
It is worth noting that here again, the observables in a $\mathrm{U}(1)^n$ BF theory would be a sub-case of a $\mathrm{U}(1)^{2n}$ CS theory since 
\begin{align*}
\label{eq:First_def_expectation_value}
\langle \langle W_M (\mathbf{A},\mathbf{B}, \boldsymbol{\eta}_1,\bh_2) \rangle \rangle_{\mathrm{BF}_{\mathbf{C}}}
=& \frac{1}{\mathcal{N}_{\mathrm{BF}_{\mathbf{C}}}(M)}
\int_{\left(H^1_{\mathrm{DB}\times }(M)\right)^{n}\times \left(H^1_{\mathrm{DB}\times }(M)\right)^{n}}
\!\!\!\!\!\!\!\!\!\!\!\!\!\!\!
\mathscr{D}\mathbf{A}\mathscr{D}\mathbf{B} 
e^{2\pi i \left( S_{\mathrm{BF}_{\mathbf{C}}}\left(\mathbf{A} , \mathbf{B}\right)+\int_M \mathbf{A}\star \boldsymbol{\eta}_1 + \int_M \mathbf{B}\star \boldsymbol{\eta}_2 \right)} \\
=&\frac{1}{\mathcal{N}_{\mathrm{CS}_{\mathbf{C'}}}(M)}
\int_{\left(H^1_{\mathrm{DB}}(M)\right)^{2n}}\mathscr{D}\mathbf{A} 
e^{2\pi i \left( S_{\mathrm{CS}_{\mathbf{C'}}}\left(\mathbf{A'}\right)+\int_M \mathbf{A'}\star \boldsymbol{\eta}' \right)} ,
\end{align*}
with $\mathbf{A}' = \mathbf{A} \oplus \mathbf{B}$, $\bh' = \bh_1 \oplus \bh_2$ and $\mathbf{C'} = \begin{pmatrix}
    \mathbf{0} & \mathbf{C} \\
    \mathbf{0} & \mathbf{0}
\end{pmatrix}$.

In principle, we can compute the expectation value of any DB class $B$ simply by replacing $\bh$ with a general $B$, but for now we will restrict to expectation values of classes/distributions, associated with loops. In general, our observables could also be links, i.e., a collection of loops. This means that now we have two types of links, the surgery link (on which Dehn surgery will be performed) and the observable link that will not undergo surgery  itself but will be affected by the surgery depending on which components it was linked to. This description with two links (one observable and one for surgery) can also be seen in \cite{mattes_invariants_1993}. In $\mathrm{U}^n(1)$, as we explained earlier, the observable link will in-fact be $n$-fold, one for each copy of the gauge group. Each of these $n$ folds can still link and interact with each other. By cutting and gluing, such a link $\gamma$ inside the manifold can always be decomposed homologically into three parts:
\[
\gamma = \gamma_0 + \gamma_t + \gamma_f.
\]
These parts correspond to the free, the torsion and the topologically trivial components of the link. Furthermore, we can always cut our link in such a way that the components are simple unknots, each belonging to one of these three parts. This decomposition of $ \bgamma$ in turn,  causes a decomposition of $\bh$. 
 \[
\begin{array}{cccccccccc}
 \bgamma &=&& \tikzmark{a}\bgamma_0 &+& \tikzmark{b}\bgamma_{t} &+ &\tikzmark{c}\bgamma_f 
 \\\\
%  y=\left[\frac{a}{\tikzmark{a}b}\frac{c}{\tikzmark{b}d}\right]\\[2ex]\\
    \bh &
    =&
    \boldsymbol{\omega} \tikzmark{d} &
    + &
   % \partial_{\Sigma\tikzmark{e} } / p 
    \boldsymbol{j}_\Sigma \tikzmark{e}
    &
    +
     &
     \bht\tikzmark{f} &
     + &
     \bh_\mathbf{f} \tikzmark{g}\\&
     \multicolumn{1}{c}{} & \multicolumn{3}{c}{$\upbracefill$}&
     \multicolumn{1}{c}{} & \multicolumn{3}{c}{$\upbracefill$}&\\&
  \multicolumn{1}{c}{} & \multicolumn{3}{c}{\text{Perturbative}}& 
  \multicolumn{1}{c}{} & \multicolumn{3}{c}{\text{Non-perturb}}\\
  &
  \multicolumn{1}{c}{} & \multicolumn{3}{c}{\text{part}}& 
  \multicolumn{1}{c}{} & \multicolumn{3}{c}{\text{part}}\\
  \tikz[remember picture,overlay]{
    \draw[->] (a.south)++(.13em,-.5ex) to ([shift={(0ex,2ex)}]d.north) ;
    \draw[->] (b.south)++(.30em,-.6ex) to ([shift={(0ex,2ex)}]e.north) ;
    \draw[->] (b.south)++(.30em,-.6ex) to ([shift={(-2.2ex,1.7ex)}]f);
    \draw[->] (c.south)++(.45em,-.7ex) to ([shift={(-2.2ex,1.4ex)}]g.north) ;
    %\draw [->] (a) -- (destination+ (1,1));
  }  
\end{array} 
\]
Let us examine the part of $\bh$ that corresponds to torsion (i.e. $ \boldsymbol{j}_\Sigma \tikzmark{e}
    +
     \bht$). The DB-tuple of a generator of order $p$ of that part would look like this:
\begin{center}
\begin{tikzpicture}
\node at (0,0) {$ (\partial_\Sigma / p , m/p , n). $};
%$(\partial_\Sigma / p , m/p , n).$
\draw[color = red]    (-.71,0. ) ellipse (5.5mm and 5.5mm);
\draw[color = red]    (0.48,-0.03) ellipse (6mm and 3mm);
 \node (3) at (-1.5,-1.5) {
        $%\boldsymbol
        {j}_\Sigma$
        };
  \node (4) at (1.5,-1.5) {
        $\eta_t$
        };
        \draw [->,black] (3) to (-1,-0.55) node[label , yshift=+3mm , xshift = -12mm  ]{};
        \draw [->,black] (4) to (0.8,-0.25) node[label , yshift=+3mm , xshift = -12mm  ]{};
\end{tikzpicture}
\end{center}
Where $j_\Sigma = \partial_\Sigma /p$ are currents, such that $p j_\Sigma$ is dual to the boundary of a surface. $\bht$ are in fact, the pseudo-canonical origins of DB \cite{guadagnini_path-integral_2014,kim_u1n_2025}. We can now perform computations with $\bh$. The field ${A}$ also admits a decomposition \cite{guadagnini_three-manifold_2013}
\begin{align*}
\label{decompA}
A = \underset{\underset{FH^{2}(M)}{\rotatebox[origin=c]{-90}{$\in$}}}{A_{{m}_A}} 
+ \underset{\underset{TH^{2}(M)}{\rotatebox[origin=c]{-90}{$\in$}}}{A_{{\kappa}_A}} 
+ \underset{\underset{\slfrac{\Omega^{1}(M)}{\Omega^{1}_{\mathrm{cl}}(M)}}
{\rotatebox[origin=c]{-90}{$\in$}}}{\alpha_{\perp}} 
+ \underset{\underset{\slfrac{\Omega^{1}_{\mathrm{cl}}(M)}{\Omega^{1}_{\Z}(M)}}
{\rotatebox[origin=c]{-90}{$\in$}}}{\alpha_{0}},
\end{align*}
which, in turn, induces a decomposition on $\mathbf{A}$. With that in mind, equation $\eqref{eq:First_def_expectation_value}$ decomposes as:
\begin{align*}
\mbox{``} \langle \langle W_M (\mathbf{A}, \boldsymbol{\eta}) \rangle \rangle_{\mathrm{CS}_{\mathbf{C}}}
= &\frac{1}{\mathcal{N}_{\mathrm{CS}_{\mathbf{C}}}(M)}
\sum_{\boldsymbol{\kappa}_{\mathbf{A}}\in\left(TH^{2}(M)\right)^{n}}
e^{2\pi i\int_{M}\left(\mathbf{A}_{\boldsymbol{\kappa}_{\mathbf{A}}}^\top \star\mathbf{C}\mathbf{A}_{\boldsymbol{\kappa}_{\mathbf{A}}}+\mathbf{A}_{\boldsymbol{\kappa}_{\mathbf{A}}}^\top \star \boldsymbol{\eta}\right)}\\
&\sum_{\mathbf{m}_{\mathbf{A}}\in\left(FH^{2}(M)\right)^{n}}e^{2\pi i\left(\int_{M}\mathbf{A}_{\boldsymbol{\kappa}_{\mathbf{A}}}^\top \star\mathbf{C}\mathbf{A}_{\mathbf{m}_{\mathbf{A}}}+ \int_{M}\mathbf{A}_{\mathbf{m}_{\mathbf{A}}}^\top \star\mathbf{C}\mathbf{A}_{\boldsymbol{\kappa}_{\mathbf{A}}} +\int_M\mathbf{A}_{\mathbf{m}_{\mathbf{A}}}^\top \star \boldsymbol{\eta}\right)}
\\
&\int_{\left(\slfrac{\Omega^{1}_{\mathrm{cl}}(M)}{\Omega^{1}_{\Z}(M)}\right)^{n}}
\,\mathscr{D}\boldsymbol{\alpha}_{0}\,
e^{2\pi i\left(\int_{M}\mathbf{A}_{\mathbf{m}_{\mathbf{A}}}^\top \star\mathbf{C}\boldsymbol{\alpha}_{0}
+ \int_{M}\boldsymbol{\alpha}_{0}^\top \star\mathbf{C}\mathbf{A}_{\mathbf{m}_{\mathbf{A}}} +\int_M\boldsymbol{\alpha}_{0}^\top \star \boldsymbol{\eta} \right)}\\
&\int_{\left(\slfrac{\Omega^{1}(M)}{\Omega^{1}_{\mathrm{cl}}(M)}\right)^{n}}
\,\mathscr{D}\boldsymbol{\alpha}_{\perp}\,
e^{2\pi i\left(\int_{M}\mathbf{A}_{\mathbf{m}_{\mathbf{A}}}^\top \star\mathbf{C}\boldsymbol{\alpha}_{\perp} 
+ \int_{M}(\boldsymbol{\alpha}_{\perp})^\top \star\mathbf{C}\mathbf{A}_{\mathbf{m}_{\mathbf{A}}} 
+ \int_{M}(\boldsymbol{\alpha}_{\perp})^\top \star\mathbf{C}\boldsymbol{\alpha}_{\perp}+\int_M(\boldsymbol{\alpha}_{\perp})^\top \star \boldsymbol{\eta}  \right)}
\mbox{''}
\end{align*}
Because distributional DB classes have integer linking \cite{guadagnini_path-integral_2014}:
\[
\int_M\mathbf{A}_{\mathbf{m}_{\mathbf{A}}}^\top \star \boldsymbol{\eta} \underset{\Z}{=}  0 \,.
\]
Now using the fact that
\[
\int_{M}\mathbf{A}_{\mathbf{m}_{\mathbf{A}}}^\top \star\mathbf{C}\boldsymbol{\alpha}_{0}
\underset{\Z}{=} \mathbf{m}_{\mathbf{A}}^\top\mathbf{C}\boldsymbol{\theta}_{\mathbf{A}},
\]

\[
\int_{M}\boldsymbol{\alpha}_{0}^\top \star\mathbf{C}\mathbf{A}_{\mathbf{m}_{\mathbf{A}}}
\underset{\Z}{=} \boldsymbol{\theta}_{\mathbf{A}}^\top\mathbf{C}\mathbf{m}_{\mathbf{A}}
\]
and
\[
\int_M\boldsymbol{\alpha}_{0}^\top \star \boldsymbol{\eta}=\int_M\boldsymbol{\alpha}_{0}^\top \star\bh_{\boldsymbol{f}} 
\underset{\Z}{=}  \boldsymbol{\theta}_{\mathbf{A}}^\top\mathbf{f}
\]
\begin{align*}
& \int_{\left(\slfrac{\R}{\Z}\right)^{n}}
\,d\boldsymbol{\theta}_{\mathbf{A}}\,
e^{2\pi i\left(\boldsymbol{\theta}_{\mathbf{A}}^\top \left( \mathbf{K}\mathbf{m}_{\mathbf{A}}+\mathbf{f}\right)\right)}\\=
    & \; \delta_{\left( \mathbf{K}\mathbf{m}_{\mathbf{A}}+\mathbf{f}\right)}.
\end{align*}
Where $\mathbf{f} \in (FH^2(M))^n \simeq (\Z^{b_1})^n$ is the class of the free homology component correspoding to $\bh_{\boldsymbol{f}}$.  As long as  the equation $\left( \mathbf{K}\mathbf{m}_{\mathbf{A}}+\mathbf{f}\right) = 0$ has a solution for integer vectors $\mathbf{f}$ and $\mathbf{m}_{\mathbf{A}}$,
the terms
$
\int_{M}\mathbf{A}_{\boldsymbol{\kappa}_{\mathbf{A}}}^\top \star\mathbf{C}\mathbf{A}_{\mathbf{m}_{\mathbf{A}}}+ \int_{M}\mathbf{A}_{\mathbf{m}_{\mathbf{A}}}^\top \star\mathbf{C}\mathbf{A}_{\boldsymbol{\kappa}_{\mathbf{A}}}
$
 will cancel out with the free part, 
$
\int_M\mathbf{A}_{\boldsymbol{\kappa}_{\mathbf{A}}}^\top \star \boldsymbol{\eta}_\mathbf{f},
$
of the term
$
\int_M\mathbf{A}_{\boldsymbol{\kappa}_{\mathbf{A}}}^\top \star \boldsymbol{\eta}
$
. The same is true for
$
\int_{M}\mathbf{A}_{\mathbf{m}_{\mathbf{A}}}^\top \star\mathbf{C}\boldsymbol{\alpha}_{\perp} 
+
\int_{M}(\boldsymbol{\alpha}_{\perp})^\top \star\mathbf{C}\mathbf{A}_{\mathbf{m}_{\mathbf{A}}} 
$, and the free part of
$
\int_M(\boldsymbol{\alpha}_{\perp})^\top \star \boldsymbol{\eta}.
$
Since we used the pseudo-canonical origins for $\mathbf{A}_{\boldsymbol{\kappa}_{\mathbf{A}}}$, then \[ \int_M\mathbf{A}_{\boldsymbol{\kappa}_{\mathbf{A}}}^\top \star \mathbf{C}(\boldsymbol{\omega} +
    \boldsymbol{j}_\Sigma)=0,
    \]
(trivial by performing the computation). For the same reason, 
\[
\int_M(\boldsymbol{\alpha}_{\perp})^\top\star \bht =0,
\]
as well. Hence, the only parts of $\bh$ coupling to $\boldsymbol{\alpha}_{\perp}$ are the perturbative parts $\boldsymbol{\omega}$ and $
    \boldsymbol{j}_\Sigma$.
  %Lastly, the support of the the trivial part $\bh_{\boldsymbol{\tau}}$ can be chosen such that it decouples from everything else.
     So we remain with:

\begin{align*}
\mbox{``}
\langle \langle W_M (\mathbf{A}, \boldsymbol{\eta}) \rangle \rangle_{\mathrm{CS}_{\mathbf{C}}}
= &\frac{1}{\mathcal{N}_{\mathrm{CS}_{\mathbf{C}}}(M)}
\sum_{\boldsymbol{\kappa}_{\mathbf{A}}\in\left(TH^{2}(M)\right)^{n}}
e^{2\pi i\int_{M}\left(\mathbf{A}_{\boldsymbol{\kappa}_{\mathbf{A}}}^\top \star\mathbf{C}\mathbf{A}_{\boldsymbol{\kappa}_{\mathbf{A}}}+\mathbf{A}_{\boldsymbol{\kappa}_{\mathbf{A}}}^\top \star \boldsymbol{\eta}_\mathbf{t}\right)}
\\
&\int_{\left(\slfrac{\Omega^{1}(M)}{\Omega^{1}_{\mathrm{cl}}(M)}\right)^{n}}
\,\mathscr{D}\boldsymbol{\alpha}_{\perp}\,
e^{2\pi i\left(\int_{M}(\boldsymbol{\alpha}_{\perp})^\top \star\mathbf{C}\boldsymbol{\alpha}_{\perp}
+\int_M(\boldsymbol{\alpha}_{\perp})^\top \star (\boldsymbol{\omega} +
    \boldsymbol{j}_\Sigma)  \right)}
\mbox{''}
\end{align*}
For the sake of notation, we will refer to $\boldsymbol{\omega} +
    \boldsymbol{j}_\Sigma$ as 
    $\bh_{\boldsymbol{p}}$
    for the next calculation ($p$ stands for perturbative). The normalization $\mathcal{N}_{\mathrm{CS}_{\mathbf{C}}}(M)$ will be defined in the same way as the partition function and will be:
    \[\mathcal{N}_{\mathrm{CS}_{\mathbf{C}}}(M) = 
\int_{\left(\slfrac{\Omega^{1}(M)}{\Omega^{1}_{\mathrm{cl}}(M)}\right)^{n}}
\,\mathscr{D}\boldsymbol{\alpha}_{\perp}\,
e^{2\pi i\left(\int_{M}(\boldsymbol{\alpha}_{\perp})^\top \star\mathbf{C}\boldsymbol{\alpha}_{\perp}
 \right)} .
    \]
    
\begin{figure}
     \centering
\begin{subfigure}[b]{0.4\textwidth}
         \centering         
         \includegraphics[width=\textwidth]{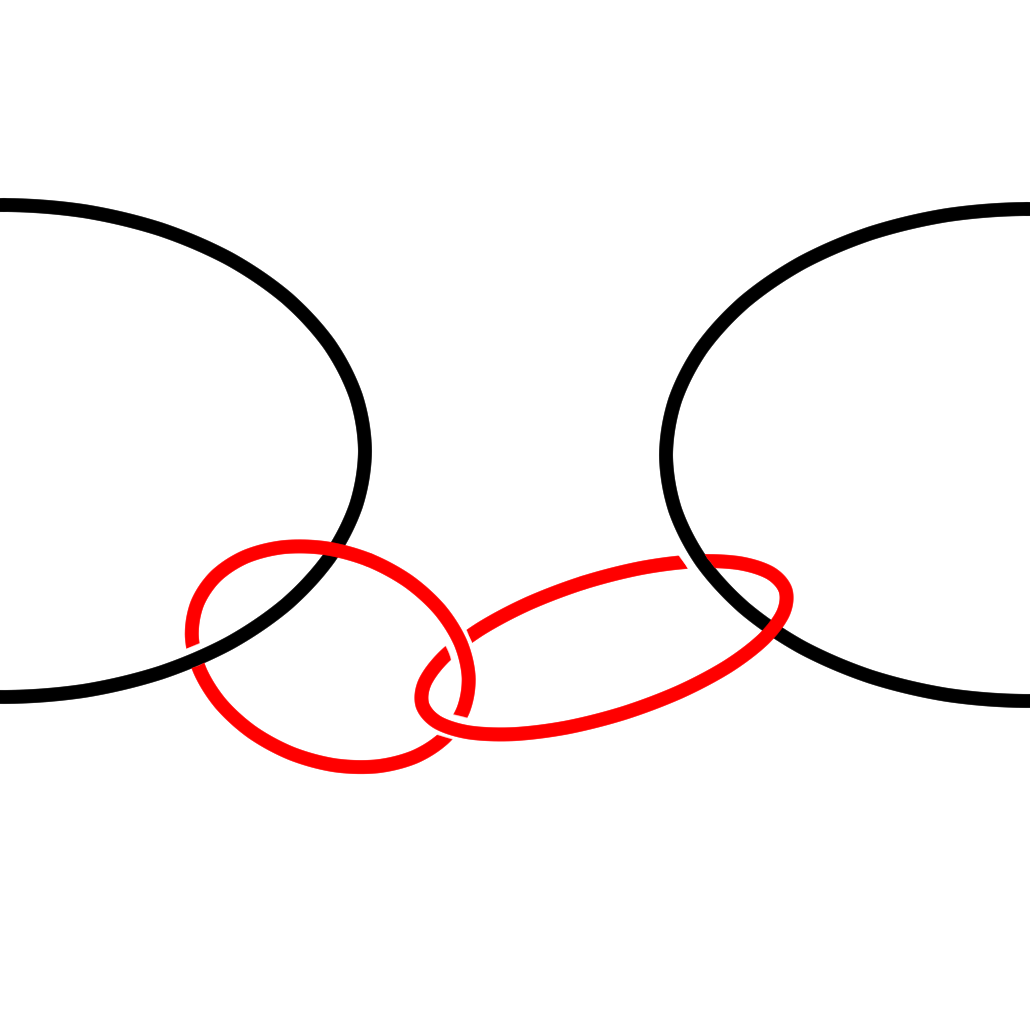}
         \caption{Two surgery components (in black) that have torsion observable components (in red) linked to them and interlinked to each other.}
         \label{Fig:interlinked torsions}
     \end{subfigure}
     \hfill
\begin{subfigure}[b]{0.4\textwidth}
         \centering         
         \includegraphics[width=1\textwidth , trim=15 0cm -15 0]{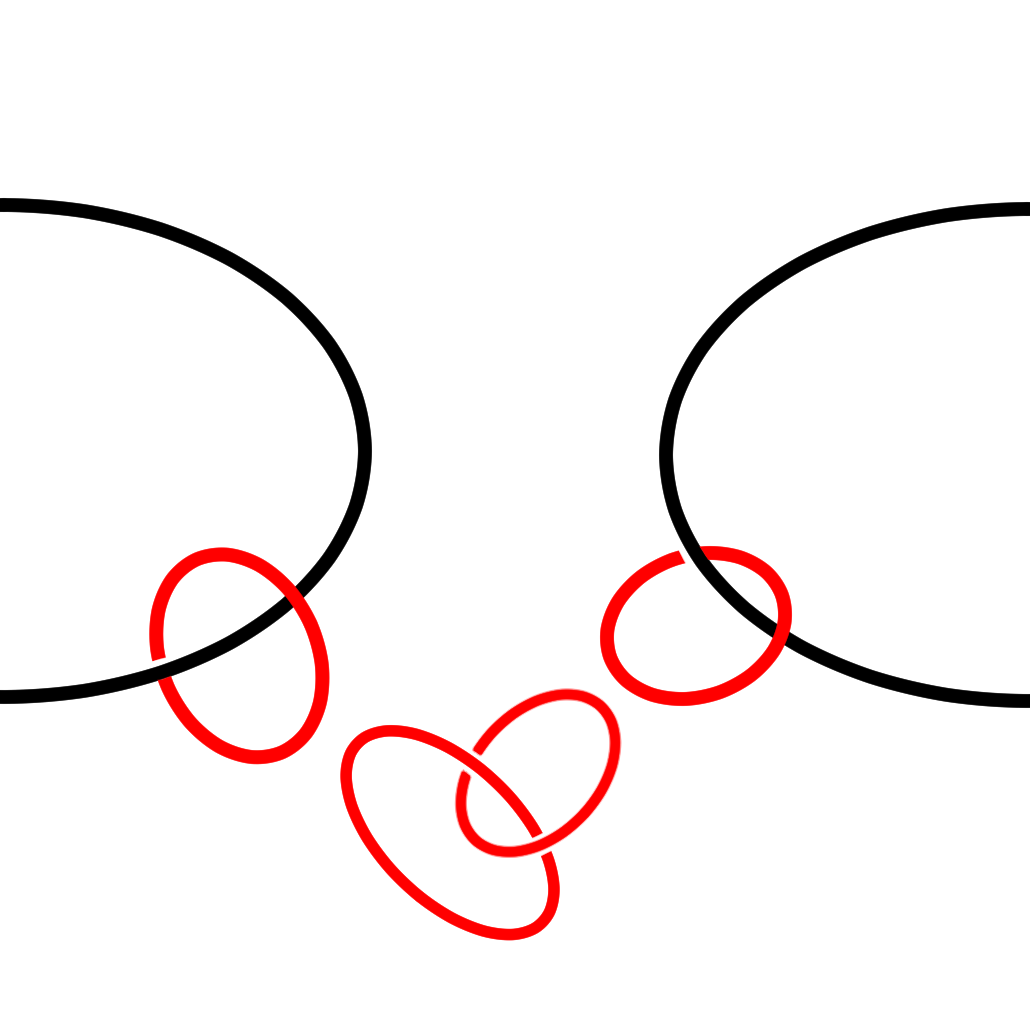}
         \caption{The same links as figure \ref{Fig:interlinked torsions} but now we have split the torsion components so that the interlinking has passed to trivial components.}
         \label{Fig:trivial split from torsion}
\end{subfigure}

\begin{subfigure}[b]{0.4\textwidth}
         \centering         
         \includegraphics[width=1\textwidth]{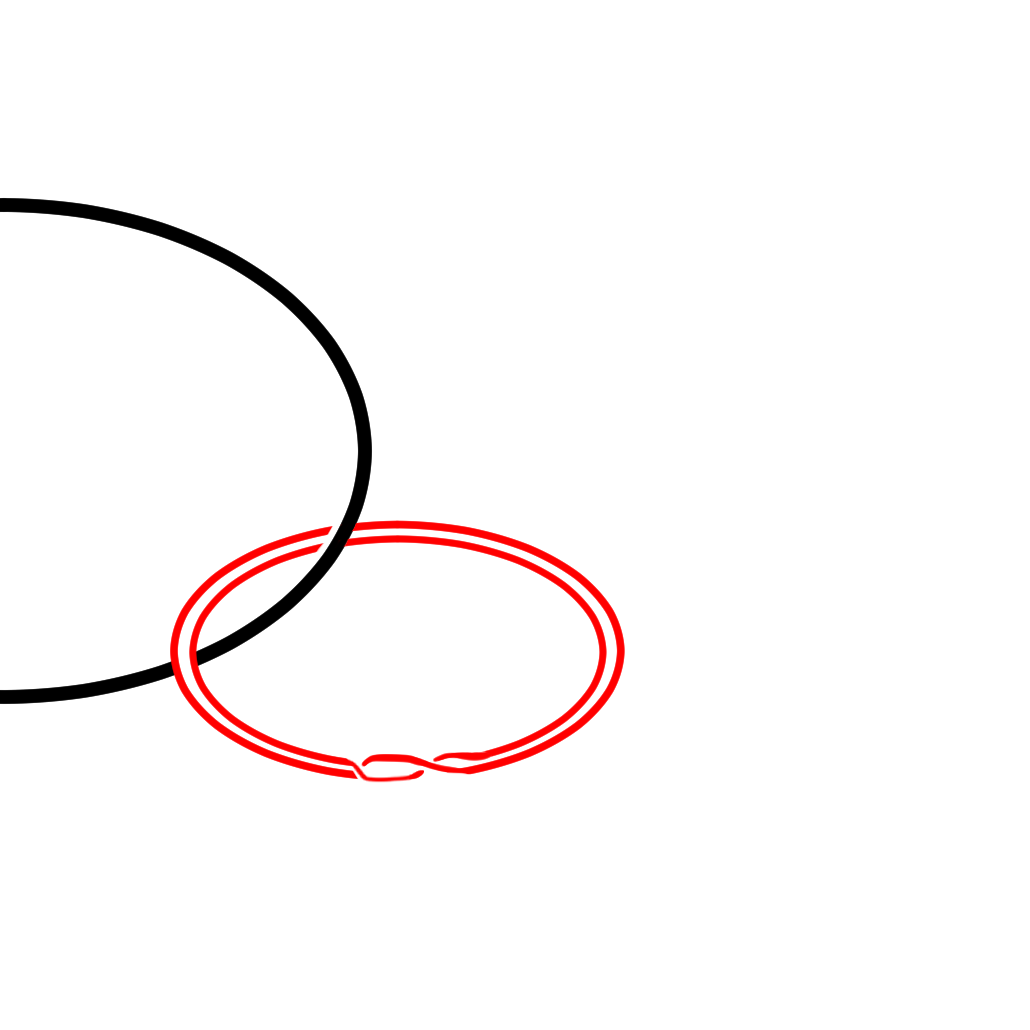}
         \caption{A surgery component that is linked to a torsion observable that has some framing (self-linking). The observable is represented as a ribbon so the framing can be shown as a twist of that ribbon.}
         \label{Fig:torsion with framing}
\end{subfigure}
\hfill
\begin{subfigure}[b]{0.4\textwidth}
         \centering         
         \includegraphics[width=1\textwidth, trim=15 0cm -15 0]{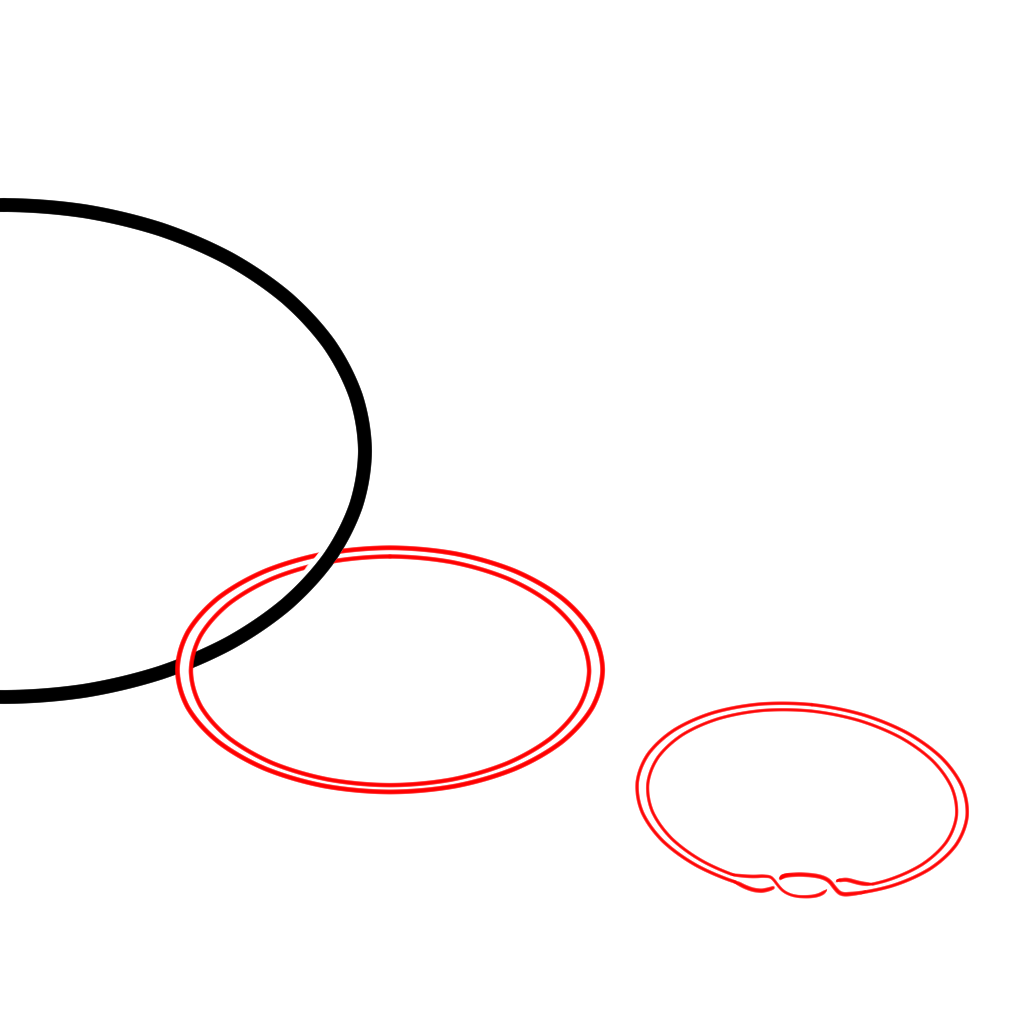}
         \caption{The same links as in figure \ref{Fig:torsion with framing} but now we have extracted the twist (i.e. the framing) into a trivial component, so the torsion observable component has zero framing.}
         \label{Fig:framing split from torsion}
\end{subfigure}
\caption{Extracting the framing and the interlinking from the torsion parts of the observable.}
\label{Fig:extracting framing}
\end{figure}

    Substituting the normalization we will have:
\begin{align*}
\langle \langle W_M (\mathbf{A}, \boldsymbol{\eta}) \rangle \rangle_{\mathrm{CS}_{\mathbf{C}}}
=& \sum_{\boldsymbol{\kappa}_{\mathbf{A}}\in\left(TH^{2}(M)\right)^{n}}
e^{-\pi i \boldsymbol{\kappa}_{\mathbf{A}} ^\top \left(\mathbf{K} \otimes \mathbf{Q}\right) \boldsymbol{\kappa}_{\mathbf{A}}-2\boldsymbol{\kappa}_{\mathbf{A}}^\top (\mathbf{Id}\otimes \mathbf{Q}) \boldsymbol{\tau}}
\\&
\frac{
\int_{\left(\slfrac{\Omega^{1}(M)}{\Omega^{1}_{\mathrm{cl}}(M)}\right)^{n}}
\,\mathscr{D}\boldsymbol{\alpha}_{\perp}\,
e^{2\pi i\left(\int_{M}(\boldsymbol{\alpha}_{\perp})^\top \star\mathbf{C}\boldsymbol{\alpha}_{\perp}
+\int_M(\boldsymbol{\alpha}_{\perp})^\top \star \bh_{\boldsymbol{p}} \right)}}{
\int_{\left(\slfrac{\Omega^{1}(M)}{\Omega^{1}_{\mathrm{cl}}(M)}\right)^{n}}
\,\mathscr{D}\boldsymbol{\alpha}_{\perp}\,
e^{2\pi i\left(\int_{M}(\boldsymbol{\alpha}_{\perp})^\top \star\mathbf{C}\boldsymbol{\alpha}_{\perp}
 \right)}} ,
 \end{align*}
 with $\boldsymbol{\tau} \in TH^2(M)$ the torsion class corresponding to $\bht$. The term 
 \[
 \frac{
\int_{\left(\slfrac{\Omega^{1}(M)}{\Omega^{1}_{\mathrm{cl}}(M)}\right)^{n}}
\,\mathscr{D}\boldsymbol{\alpha}_{\perp}\,
e^{2\pi i\left(\int_{M}(\boldsymbol{\alpha}_{\perp})^\top \star\mathbf{C}\boldsymbol{\alpha}_{\perp}
+\int_M(\boldsymbol{\alpha}_{\perp})^\top \star \bh_{\boldsymbol{p}} \right)}}{
\int_{\left(\slfrac{\Omega^{1}(M)}{\Omega^{1}_{\mathrm{cl}}(M)}\right)^{n}}
\,\mathscr{D}\boldsymbol{\alpha}_{\perp}\,
e^{2\pi i\left(\int_{M}(\boldsymbol{\alpha}_{\perp})^\top \star\mathbf{C}\boldsymbol{\alpha}_{\perp}
 \right)}} 
 \]
 is called the perturbative term of the expectation value. To calculate that, we will focus on the exponent of the numerator. For the next calculation, we will use the notation $\bfK^{-1}\bh_{\boldsymbol{p}}$. Rigorously, this is an abuse of notation since we cannot multiply DB classes by a rational number. It only makes sense here since the class $\bh_{\boldsymbol{p}}$ is of the form $[(\boldsymbol{a},0,0)]$ so by $\bfK^{-1}\bh_{\boldsymbol{p}}$ we mean the class $[(\bfK^{-1}\boldsymbol{a} ,0,0)]$. Before we work on the exponent, we want to perform the following calculation:
 %\[ 2\pi i \left(\int_{M}(\boldsymbol{\alpha}_{\perp})^\top \star\mathbf{C}\boldsymbol{\alpha}_{\perp}+\int_M(\boldsymbol{\alpha}_{\perp})^\top \star \bh_{\boldsymbol{p}} \right) \]
  \[
 2\pi i \int_{M}\left((\boldsymbol{\alpha}_{\perp})^\top+\bfK^{-1}\,\bh_{\boldsymbol{p}}^\top\right)\star\mathbf{C}\left(\boldsymbol{\alpha}_{\perp}+\bfK^{-1}\bh_{\boldsymbol{p}}\right) = 
 \]
 \[
 2\pi i \int_{M}\left((\boldsymbol{\alpha}_{\perp})^\top \star\mathbf{C}\boldsymbol{\alpha}_{\perp}
+
(\boldsymbol{\alpha}_{\perp})^\top \star\mathbf{C}\bfK^{-1}\bh_{\boldsymbol{p}}
+
\bfK^{-1}\,\bh_{\boldsymbol{p}}^\top \star\mathbf{C}\boldsymbol{\alpha}_{\perp}
+
\bfK^{-1}\,\bh_{\boldsymbol{p}}^\top \star\mathbf{C}\bfK^{-1}\bh_{\boldsymbol{p}}
\right)=
 \]
 \[2\pi i \int_{M}\left((\boldsymbol{\alpha}_{\perp})^\top \star\mathbf{C}\boldsymbol{\alpha}_{\perp}
+
(\boldsymbol{\alpha}_{\perp})^\top \star(\mathbf{C}+\mathbf{C}^\top)\bfK^{-1}\bh_{\boldsymbol{p}}
+
\bfK^{-1}\,\bh_{\boldsymbol{p}}^\top \star\mathbf{C}\bfK^{-1}\bh_{\boldsymbol{p}}
\right)=
 \]
   \[
 2\pi i \int_{M}\left((\boldsymbol{\alpha}_{\perp})^\top \star\mathbf{C}\boldsymbol{\alpha}_{\perp}
+
(\boldsymbol{\alpha}_{\perp})^\top \star\bh_{\boldsymbol{p}}
+
\bfK^{-1}\,\bh_{\boldsymbol{p}}^\top \star\mathbf{C}\bfK^{-1}\bh_{\boldsymbol{p}}
\right)
 \]
 So now we can rewrite the exponent in the numerator as
 \[
  2\pi i \int_{M}\left((\boldsymbol{\alpha}_{\perp})^\top+\bfK^{-1}\,\bh_{\boldsymbol{p}}^\top\right)\star\mathbf{C}\left(\boldsymbol{\alpha}_{\perp}+\bfK^{-1}\bh_{\boldsymbol{p}}\right) -2\pi i \int_{M}\bfK^{-1}\,\bh_{\boldsymbol{p}}^\top \star\mathbf{C}\bfK^{-1}\bh_{\boldsymbol{p}}
 \]
 The first term is a shift on the infinite dimensional part which gets eliminated by the normalization. For the second term we can do the following:
  \[
  -2\pi i \int_{M}\bfK^{-1}\,\bh_{\boldsymbol{p}}^\top \star\mathbf{C}\bfK^{-1}\bh_{\boldsymbol{p}}=   -2\pi i \int_{M}\bfK^{-1}\,\bh_{\boldsymbol{p}}^\top \wedge\mathbf{C}\bfK^{-1}d\bh_{\boldsymbol{p}}= \]
  \[
  -\pi i \int_{M}\left(\bfK^{-1}\,\bh_{\boldsymbol{p}}^\top \wedge\mathbf{C}\bfK^{-1}d\bh_{\boldsymbol{p}}+\bfK^{-1}\,\bh_{\boldsymbol{p}}^\top \wedge\mathbf{C}^\top\bfK^{-1}d\bh_{\boldsymbol{p}} \right) = 
  -\pi i \int_{M}\bfK^{-1}\,\bh_{\boldsymbol{p}}^\top \wedge\mathbf{K}\bfK^{-1}d\bh_{\boldsymbol{p}}= 
 \]
There would be three contributions to this term:
 \[
 \bfK^{-1}\,\boldsymbol{\omega}^\top \wedge\mathbf{K}\bfK^{-1}d\boldsymbol{\omega}\text{,   }\hspace{0.3cm} \bfK^{-1}\,\boldsymbol{j}_\Sigma^\top \wedge\mathbf{K}\bfK^{-1}d\boldsymbol{j}_\Sigma\hspace{0.3cm}
 \text{   and   }\hspace{0.3cm}
 2\bfK^{-1}\,\boldsymbol{\omega} ^\top \wedge\mathbf{K}\bfK^{-1}d\boldsymbol{j}_\Sigma.
 \]
 The first term is nothing but the linking and self-linking between the trivial components of the loop:
 \[
   -\pi i (\bfK^{-1} \otimes \boldsymbol{\ell} \boldsymbol{k})(
   \boldsymbol{\gamma}_{\boldsymbol{0}},\boldsymbol{\gamma}_{\boldsymbol{0}}).
 \]
To compute this, we would need a framing on the trivial parts of the loop. The last term is the linking between the torsion and trivial parts of the loop:
\[
   -\pi i (\bfK^{-1} \otimes \boldsymbol{\ell} \boldsymbol{k})(
   \boldsymbol{\gamma}_{\boldsymbol{0}},\boldsymbol{\gamma}_{\boldsymbol{t}}).
 \]
Note that we can always take this term to be $0$, again by cutting and gluing all torsion generators to be unlinked with the rest of the link. An example can be seen later in figure \ref{Fig:extracting framing}. Now, the middle term requires a small computation. Because of zero regularization we have: 
\[
\int_M (\bht+ \boldsymbol{j}_\Sigma)^\top \star C(\bht+ \boldsymbol{j}_\Sigma) \stackrel{\mod \Z}{=} 0. 
\]
But also,
\[
\int_M (\bht+ \boldsymbol{j}_\Sigma)^\top \star C(\bht+ \boldsymbol{j}_\Sigma) =\int_M \bht ^\top \star C\bht+\int_M \boldsymbol{j}_\Sigma ^\top \star C \boldsymbol{j}_\Sigma,
\]
and
\[
\int_M \bht ^\top \star C\bht \stackrel{\mod \Z}{=}  - (\mathbf{C} \otimes \bfQ)(\boldsymbol{\tau},\boldsymbol{\tau}).
\]
This means that:
\begin{equation}
    \label{eq:zeroregform}
    \int_M \boldsymbol{j}_\Sigma ^\top \star C \boldsymbol{j}_\Sigma \stackrel{\mod \Z}{=} (\mathbf{C} \otimes \bfQ)(\boldsymbol{\tau},\boldsymbol{\tau}).
\end{equation}
Here though we have something different, we had 
\[
-2\pi i \int_{M}\bfK^{-1}\,\boldsymbol{j}_\Sigma^\top \wedge\mathbf{C}\bfK^{-1}d\boldsymbol{j}_\Sigma, 
\]
or equivalently by the commutativity of wedge product:
\[
-\pi i \int_{M}\bfK^{-1}\,\boldsymbol{j}_\Sigma^\top \wedge\mathbf{K}\bfK^{-1}d\boldsymbol{j}_\Sigma. 
\]
Assigning a value to this quantity is a bit tricky. The problem is that now, we cannot work $\mod \Z$ as before since our currents are multiplied by a rational matrix. To be able to assign a meaningful value to this  quantity, we need to use a geometric point of view. We will use the fact that our manifold can be constructed by Dehn surgery, on a link $\mathcal{L}$ described by the linking matrix $\bfL$. Afterwards we will consider how the loop $\boldsymbol{\gamma}_{\boldsymbol{t}}$ intersects $\bfL$ before the surgery, while living in $S^3$. Note that up until now we did not require our manifold to be coming from Dehn surgery. As mentioned in the introduction, this is still true but this picture allows us a consistent way to choose representatives for the previous quantities. For simplicity we will take $\bfL$ to be non-degenerate, but we will examine the degenerate case later too.
\\
\\
Given a surgery link $\mathcal{L}$ and our observable loop $\boldsymbol{\gamma}|_{S^3}$, now in $S^3$ before the surgery, we can use the following ideas. First of all, we do not care about the part of the observable corresponding (after surgery) to the free sector as this will vanish. Secondly, without loss of generality, we can consider the framing (self-linking) of the torsion part of the observable loop to be $0$ in $S^3$. That is because we can always transfer this framing to a new trivial component that we can split from the torsion part \cite{guadagnini_path-integral_2014}. Lastly, we can consider the torsion parts to be unlinked to each other, that is because, similar to the previous idea, we can always extract this linking to a trivial component. All of these ideas can be seen below in figure \ref{Fig:extracting framing}. This means that we can always take the torsion parts to be unknots that link simply to each component of the surgery link. This is a consequence of the theory being abelian, all the expectation values are based on linking numbers and thus we can cut and glue as long as we keep the linking intact \cite{guadagnini_path-integral_2014,guadagnini_three-manifold_2013}. In the cases where they are linked twice or more, we will say that the observable has a charge equal to that linking number. Note that just because the torsion parts are not linked to each other in $S^3$ (pre-surgery), does not mean that they do not interact. In fact, the reason we want them to be unlinked is to avoid extra interaction terms. These components still interact naturally through the linking form (for observable components coupled to different surgery components) and through the $\bfK$ matrix (for observable components belonging in different copies of the gauge group). Now we can start the computation. 
\\
\\
We could represent $\bgt$ by the vector $\boldsymbol{\ell}$ whose components can be defined as $\boldsymbol{\ell}_i = \ell k(\bgt, \mathcal{L}_i)|_{S^3}$. This $\boldsymbol{\ell}$ is closely related to $\boldsymbol{\tau}$, in fact, $\boldsymbol{\ell}$ is a specific representative of the class of $\boldsymbol{\tau}$ expressed in the basis of $\left(\slfrac{\Z^{m}}{\mathbf{L}\Z^{m}}\right)^{n}$ which is isomorphic to $TH^2(M)$. In other words, the class of $\boldsymbol{\ell}$ is equivalent to $\boldsymbol{\tau}$ under the above isomorphism. Now we want to define $\ell k(\bgt,\bgt)|_{M}$ but in $\Q$ instead of $\Q/\Z$. By the properties of the linking matrix, the loop  $(\bfL\otimes Id)\bgt$ would be a trivial loop after surgery (i.e., the boundary of a surface). Thus, the linking number $\ell k(\bgt,(\bfL\otimes Id)\bgt)|_{M}$ would be an integer and equal to $\boldsymbol{\ell}^T \cdot \boldsymbol{\ell}$. So now we can set
\[
\ell k(\bgt,\bgt)|_{M} = \boldsymbol{\ell}^T(\bfL^{-1}\otimes \mathbf{Id})\boldsymbol{\ell}
\]
This now gives us a meaningful way to assign a value to the expression as
\[
 -\pi i \int_{M}\bfK^{-1}\,\boldsymbol{j}_\Sigma\wedge\mathbf{K}\bfK^{-1}d\boldsymbol{j}_\Sigma {=} -\pi i \ell k^\top (\bgt , (\mathbf{Id} \otimes \bfK ^{-1}) \bgt) =   -\pi i (\bfK^{-1} \otimes \bfL^{-1})(\boldsymbol{\ell} , \boldsymbol{\ell}).
\]

%Equation \ref{eq:zeroregform} is true only up to integers, which means that in principle
%\[
 %-\pi i \int_{M}\bfK^{-1}\,\boldsymbol{j}_\Sigma\wedge\mathbf{K}\bfK^{-1}d\boldsymbol{j}_\Sigma {=}  -\pi i ^\top (\bfK^{-1} \otimes \bfQ)(\bht , \bht)+ N^\top\bfK^{-1}N,
%\]
%With $N$ some integer vector.

And since $\bfQ(\boldsymbol{\tau},\boldsymbol{\tau}) \stackrel{\mod \Z}{=} \bfL^{-1}(\boldsymbol{\ell},\boldsymbol{\ell})$, we can write the whole thing as:
\begin{align}
\label{eq:final_obs}
\langle \langle W_M (\mathcal{L}, \boldsymbol{\gamma}|_{S^3}) \rangle \rangle_{\mathrm{CS}_{\mathbf{K}}}
=& \sum_{\boldsymbol{\kappa}_{\mathbf{A}}\in\left(\slfrac{\Z^{m}}{\mathbf{L}\Z^{m}}\right)^{n}}
e^{-\pi i \boldsymbol{\kappa}_{\mathbf{A}} ^\top \left(\mathbf{K} \otimes \mathbf{L}^{-1}\right) \boldsymbol{\kappa}_{\mathbf{A}}-2\boldsymbol{\kappa}_{\mathbf{A}}^\top (\mathbf{Id}\otimes \mathbf{L}^{-1}) \boldsymbol{\ell}}
\\&\nonumber 
e^{-\pi i (\bfK^{-1} \otimes \boldsymbol{\ell} \boldsymbol{k})(
   \boldsymbol{\gamma}_{\boldsymbol{0}},\boldsymbol{\gamma}_{\boldsymbol{0}}+\boldsymbol{\gamma}_{\boldsymbol{t}})}
   e^{ -\pi i (\bfK^{-1} \otimes \bfL^{-1})(\boldsymbol{\ell} , \boldsymbol{\ell})},
 \end{align}
where now we replaced the manifold with its surgery link $\mathcal{L}$ and the observable has been replaced with an observable link pre-surgery. Moreover, we label the expectation with the subscript ${\mathrm{CS}_{\mathbf{K}}}$ instead of ${\mathrm{CS}_{\mathbf{C}}}$ because now it is based on the matrix $\mathbf{K}$. 
 In the next section we will analyze the remaining term and also see the cases where $\bfK$ and $\bfL$ are degenerate. Note that the degeneracy of $\bfL$ is related to the free homology of the manifold. This means that everything involving $\bh_{\mathbf{f}}$, including the condition $\left( \mathbf{K}\mathbf{m}_{\mathbf{A}}+\mathbf{f}\right) = 0$, are relevant only when $\bfL$ is degenerate.
 \subsection{Reciprocity and degenerate case}
 In previous articles, we dealt with terms of the form:
 \[
\sum_{\boldsymbol{\kappa}_{\mathbf{A}}\in\left(TH^{2}(M)\right)^{n}}
e^{-\pi i \boldsymbol{\kappa}_{\mathbf{A}} ^\top \left(\mathbf{K} \otimes \mathbf{Q}\right) \boldsymbol{\kappa}_{\mathbf{A}}}
 \]
 through the use of a reciprocity formula. Now we are dealing with a term of similar form, albeit with a linear component as well. Thankfully, for this case there is also a reciprocity formula but we have to adapt the parameters. For a non-degenerate symmetric  $n\times n$ matrix $\bfK$ and a non-degenerate symmetric  $m \times m $ matrix $\bfL$, from \cite{deloup_reciprocity_2007}, we know that the following reciprocity formula is true:

\begin{align}
\label{eq:rec3} 
\nonumber
&\frac{1}{\abs{\determinant{\mathbf{K}}}^{\frac{m}{2}}}
\sum\limits_{\mathbf{x}_{\boldsymbol{\mathcal{L}}}\in\left(\slfrac{\Z^{n}}{\mathbf{K}\Z^{n}}\right)^{m}}
e^{\pi i\mathbf{x}_{\boldsymbol{\mathcal{L}}}
^\top \left(\mathbf{L}\otimes\mathbf{K}^{-1}\right)
\mathbf{x}_{\boldsymbol{\mathcal{L}}}-\pi i\mathbf{x}_{\boldsymbol{\mathcal{L}}} 
^\top \left(\mathbf{L}\otimes\mathbf{Id}_{n}\right)\mathbf{u}}\\ 
&=\frac{1}{\abs{\determinant{\mathbf{L}}}^{\frac{n}{2}}}
e^{\frac{i\pi}{4}\left(\sigma\left(\mathbf{K}\right)\sigma\left(\mathbf{L}\right)-\mathbf{u}^\top\mathbf{e}\mathbf{u}\right)}
\sum_{\boldsymbol{\kappa}_{\mathbf{A}}\in\left(\slfrac{\Z^{m}}{\mathbf{L}\Z^{m}}\right)^{n}}
e^{-\pi i \boldsymbol{\kappa}_{\mathbf{A}}
^\top \left(\mathbf{K}\otimes\mathbf{L}^{-1}\right)
\boldsymbol{\kappa}_{\mathbf{A}}+\pi i\boldsymbol{\kappa}_{\mathbf{A}}
^\top \left(\mathbf{K}\otimes\mathbf{Id}_{m}\right)\mathbf{u}'}.  
\end{align}
Where $\mathbf{u}$ is a rational Wu class for $\mathbf{e}=\mathbf{L}\otimes\mathbf{K}$, i.e., it obeys the equation $\mathbf{x}^\top \mathbf{e} \mathbf{x} + \mathbf{x}^\top \mathbf{e}\mathbf{u} \in 2\Z$ for all $\mathbf{x} \in \Z^{mn}$. 
The primed variables (e.g., $\mathbf{u}'$) signify the images of their non-primed counterparts under tensor permutation, meaning that $\mathbf{x}^\top  \left( \mathbf{L}\otimes\mathbf{K} \right)\mathbf{y} = \mathbf{x}'^\top  \left( \mathbf{K}\otimes\mathbf{L} \right)\mathbf{y}'$. For this notation, it is easy to see that $\mathbf{x}'' = \mathbf{x}$.
Now, if one of the matrices $\mathbf{L}$ or $\mathbf{K}$ is even, we know that $ 2(\mathbf{L}\otimes\mathbf{K})^{-1}\mathbf{w}$ with $\mathbf{w}$ an integer vector, will always be such a rational Wu class ($\mathbf{x}^\top \mathbf{e} \mathbf{x} \in 2\Z$ by the evenness and $\mathbf{x}^\top \mathbf{e}\mathbf{u} = 2\mathbf{x}^\top \mathbf{w} \in 2\Z$). By choosing $\mathbf{w} = -\boldsymbol{\ell}'$ and substituting $\mathbf{u}' = 2(\mathbf{K}\otimes\mathbf{L})^{-1}\boldsymbol{\ell}$ and  $ \mathbf{u} = 2(\mathbf{L}\otimes\mathbf{K})^{-1}\boldsymbol{\ell}'$ we get:

\begin{align}
\label{eq:lin_observ} 
&\frac{1}{\abs{\determinant{\mathbf{K}}}^{\frac{m}{2}}}
\sum\limits_{\mathbf{x}_{\boldsymbol{\mathcal{L}}}\in\left(\slfrac{\Z^{n}}{\mathbf{K}\Z^{n}}\right)^{m}}
e^{\pi i\mathbf{x}_{\boldsymbol{\mathcal{L}}}
^\top \left(\mathbf{L}\otimes\mathbf{K}^{-1}\right)
\mathbf{x}_{\boldsymbol{\mathcal{L}}}+2\pi i\mathbf{x}_{\boldsymbol{\mathcal{L}}} 
^\top \left(\mathbf{Id}_m\otimes\mathbf{K}^{-1}\right)\boldsymbol{\ell}'}\\ 
\nonumber
&=\frac{1}{\abs{\determinant{\mathbf{L}}}^{\frac{n}{2}}}
e^{\frac{i\pi}{4}\left(\sigma\left(\mathbf{K}\right)\sigma\left(\mathbf{L}\right)-4\boldsymbol{\ell}'^\top (\bfL \otimes \bfK)^{-1}\boldsymbol{\ell}'\right)}
\sum_{\boldsymbol{\kappa}_{\mathbf{A}}\in\left(\slfrac{\Z^{m}}{\mathbf{L}\Z^{m}}\right)^{n}}
e^{-\pi i \boldsymbol{\kappa}_{\mathbf{A}}
^\top \left(\mathbf{K}\otimes\mathbf{L}^{-1}\right)
\boldsymbol{\kappa}_{\mathbf{A}}-2\pi i\boldsymbol{\kappa}_{\mathbf{A}}
^\top \left(\mathbf{Id}_n\otimes\mathbf{L}^{-1}\right)\boldsymbol{\ell}}.  
\end{align}
So far, $\bfL$ and $\bfK$ have been non-degenerate. In general, our matrices might be degenerate so we have to account for that case. When calculating the partition function, we could compute the contribution of a $0$-sector quite easily. Now we have to address a few issues. First of all, by changing the dimension of $\bfL$ we need to change the dimension of $\mathbf{u}$. If $\bfL$ is a degenerate $m\times m$ matrix of rank $r$, it can always be transformed by Kirby moves to $\bfL \rightarrow \bfL_0 \oplus \mathbf{0}_{m-r}$ \cite{moussard_realizing_2015,tagaris_u1u1_2023} where now $\bfL_0$ is an $r\times r$ non-degenerate matrix. Now assuming that we have a $\mathbf{u}$ whose projection on the non-degenerate subspace of $\bfL$ is $\mathbf{u}_\mathrm{right}$ (the notation will become clearer later) then the sum on the left-hand-side of \ref{eq:rec3} will become:
\begin{align*}
%\label{eq:observ}
\sum\limits_{\mathbf{x}_{\boldsymbol{\mathcal{L}}}\in\left(\slfrac{\Z^{n}}{\mathbf{K}\Z^{n}}\right)^{m}}&
e^{\pi i\mathbf{x}_{\boldsymbol{\mathcal{L}}}
^\top \left(\mathbf{L}\otimes\mathbf{K}^{-1}\right)
\mathbf{x}_{\boldsymbol{\mathcal{L}}}-\pi i\mathbf{x}_{\boldsymbol{\mathcal{L}}} 
^\top \left(\mathbf{L}\otimes\mathbf{Id}_{n}\right)\mathbf{u}}= \\
\nonumber
=&\sum\limits_{\mathbf{x}_{\boldsymbol{\mathcal{L}}}\in\left(\slfrac{\Z^{n}}{\mathbf{K}\Z^{n}}\right)^{r}}
e^{\pi i\mathbf{x}_{\boldsymbol{\mathcal{L}}}
^\top \left(\mathbf{L}_0\otimes\mathbf{K}^{-1}\right)
\mathbf{x}_{\boldsymbol{\mathcal{L}}}-\pi i\mathbf{x}_{\boldsymbol{\mathcal{L}}} 
^\top \left(\mathbf{L}_0\otimes\mathbf{Id}_{n}\right)\mathbf{u}_\mathrm{right}}
\cdot \sum\limits_{\mathbf{x}_{\boldsymbol{\mathcal{L}}}\in\left(\slfrac{\Z^{n}}{\mathbf{K}\Z^{n}}\right)^{m-r}} 1  \\
=& \abs{\determinant{\mathbf{K}}}^{m-r} \sum\limits_{\mathbf{x}_{\boldsymbol{\mathcal{L}}}\in\left(\slfrac{\Z^{n}}{\mathbf{K}\Z^{n}}\right)^{r}}
e^{\pi i\mathbf{x}_{\boldsymbol{\mathcal{L}}}
^\top \left(\mathbf{L}_0\otimes\mathbf{K}^{-1}\right)
\mathbf{x}_{\boldsymbol{\mathcal{L}}}-\pi i\mathbf{x}_{\boldsymbol{\mathcal{L}}} 
^\top \left(\mathbf{L}_0\otimes\mathbf{Id}_{n}\right)\mathbf{u}_\mathrm{right}}.
\end{align*}
So, we get a factor of $\abs{\determinant{K}}$ for every degenerate direction we add. We see that the calculation is the same as for the partition function but we only use a projection of $\mathbf{u}$ at the end. The sum on the right-hand-side, will be the same as long as we use $\bfL_0$ (and its appropriate dimensions) instead of $\bfL$ and $\mathbf{u}_\mathrm{right}$. The phase will remain the same. A similar computation occurs for the right hand side but it will have a different projection $\mathbf{u}'_\mathrm{left}$, namely a projection on the space where $\bfK$ is degenerate this time. And finally, there will be one more projection, the projection of $\mathbf{u}$ onto both of these subspaces $\mathbf{u}_\mathrm{middle}$. For both $\bfK$ and $\bfL$ degenerate, we would have the following formula:
\begin{align}
\label{eq:rec5} 
\nonumber
&\frac{1}{\abs{\determinant{\mathbf{K}_{0}}}^{m-\frac{r}{2}}}
\sum\limits_{\mathbf{x}_{\boldsymbol{\mathcal{L}}}\in\left(\slfrac{\Z^{s}}{\mathbf{K}_{0}\Z^{s}}\right)^{m}}
e^{\pi i\mathbf{x}_{\boldsymbol{\mathcal{L}}}
^\top \left(\mathbf{L}\otimes\mathbf{K}_{0}^{-1}\right)
\mathbf{x}_{\boldsymbol{\mathcal{L}}}-\pi i\mathbf{x}_{\boldsymbol{\mathcal{L}}} 
^\top \left(\mathbf{L}\otimes\mathbf{Id}_{s}\right)\mathbf{u}_\mathrm{left}}\\ 
&=\frac{1}{\abs{\determinant{\mathbf{L}_{0}}}^{n-\frac{s}{2}}}
e^{\frac{i\pi}{4}\left(\sigma\left(\mathbf{K}\right)\sigma\left(\mathbf{L}\right)-\mathbf{u}^\top\mathbf{e}\mathbf{u}\right)}
\sum_{\boldsymbol{\kappa}_{\mathbf{A}}\in\left(\slfrac{\Z^{r}}{\mathbf{L}_{0}\Z^{r}}\right)^{n}}
e^{-\pi i \boldsymbol{\kappa}_{\mathbf{A}}
^\top \left(\mathbf{K}\otimes\mathbf{L}^{-1}_{0}\right)
\boldsymbol{\kappa}_{\mathbf{A}}+\pi i\boldsymbol{\kappa}_{\mathbf{A}}
^\top \left(\mathbf{K}\otimes\mathbf{Id}_{r}\right)\mathbf{u}'_\mathrm{right}}  
\end{align}
Note that we can replace $\mathbf{u}^\top\mathbf{e}\mathbf{u}$ in the phase by $\mathbf{u}_\mathrm{middle}^\top \left( \bfL_0 \otimes \bfK_0\right)\mathbf{u}_\mathrm{middle}$. It is easy to see now that the following will be a Wu class for $\mathbf{e}$, $ \mathbf{u} = 2((\mathbf{L}_0^{-1}\oplus \mathbf{0})\otimes(\mathbf{K}_0^{-1}\oplus \mathbf{0}))^{}\mathbf{w}_\mathbf{}'$, for some integer vector $\mathbf{w}$. We will choose  $\mathbf{w} = \boldsymbol{\ell}$ with $\boldsymbol{\ell}$ the vector we defined earlier from the linking of the observable loop.  Similarly we would have  $ \mathbf{u}'_\mathrm{right} = 2((\mathbf{K}_0^{-1}\oplus \mathbf{0})\otimes\mathbf{L}_0^{-1})^{}\boldsymbol{\ell}{_\mathrm{right}}_\mathbf{}$, $ \mathbf{u}_\mathrm{left} = 2((\mathbf{L}_0^{-1}\oplus \mathbf{0})\otimes\mathbf{K}_0^{-1})^{}\boldsymbol{\ell}'{_\mathrm{left}}_\mathbf{}$ and 
$\mathbf{u}_\mathrm{middle} = 2(\mathbf{L}_0\otimes\mathbf{K}_0)^{-1}\boldsymbol{\ell}'{_\mathrm{middle}}_\mathbf{}$. With those substitutions in mind the equation becomes:
\begin{align}
\label{eq:lin_recip} 
\nonumber
&\frac{1}{\abs{\determinant{\mathbf{K}_{0}}}^{m-\frac{r}{2}}}
\sum\limits_{\mathbf{x}_{\boldsymbol{\mathcal{L}}}\in\left(\slfrac{\Z^{s}}{\mathbf{K}_{0}\Z^{s}}\right)^{m}}
e^{\pi i\mathbf{x}_{\boldsymbol{\mathcal{L}}}
^\top \left(\mathbf{L}\otimes\mathbf{K}_{0}^{-1}\right)
\mathbf{x}_{\boldsymbol{\mathcal{L}}}-2\pi i\mathbf{x}_{\boldsymbol{\mathcal{L}}} 
^\top \left((\mathbf{Id}_r \oplus \mathbf{0})\otimes\bfK_0^{-1}\right)\boldsymbol{\ell}'_\mathrm{left}}\\ \nonumber
&=\frac{1}{\abs{\determinant{\mathbf{L}_{0}}}^{n-\frac{s}{2}}}
e^{\frac{i\pi}{4}\left(\sigma\left(\mathbf{K}\right)\sigma\left(\mathbf{L}\right)-4\boldsymbol{\ell}'{_\mathrm{middle}}^\top (\mathbf{L}_0\otimes\mathbf{K}_0)^{-1}\boldsymbol{\ell}'{_\mathrm{middle}}_\mathbf{}\right)}
\\&
\sum_{\boldsymbol{\kappa}_{\mathbf{A}}\in\left(\slfrac{\Z^{r}}{\mathbf{L}_{0}\Z^{r}}\right)^{n}}
e^{-\pi i \boldsymbol{\kappa}_{\mathbf{A}}
^\top \left(\mathbf{K}\otimes\mathbf{L}^{-1}_{0}\right)
\boldsymbol{\kappa}_{\mathbf{A}}+2\pi i\boldsymbol{\kappa}_{\mathbf{A}}
^\top \left((\mathbf{Id}_s \oplus \mathbf{0}) \otimes\bfL_0^{-1}\right)\boldsymbol{\ell}_\mathrm{right}}.  
\end{align}
As we mentioned earlier $\boldsymbol{\ell}_\mathrm{right}$ (or $\boldsymbol{\ell}$ when $\bfL$ is invertible) is related to what we were calling $\boldsymbol{\tau}$. While $\boldsymbol{\ell}_\mathrm{right}$ is a specific representative of a class belonging in $\left(\slfrac{\Z^{r}}{\mathbf{L}_{0}\Z^{r}}\right)^{n}$, $\boldsymbol{\tau}$ is an abstract element of $TH^2(M)$. 
%The relation is very subtle though. The homotopy class of $\bht$ is in a way a topological invariant of the theory and in this case, it belongs to the torsion group $TH^2(M)$. $\boldsymbol{\ell}_\mathrm{right}$ is $\bht$ expressed in the basis of $\left(\slfrac{\Z^{r}}{\mathbf{L}_{0}\Z^{r}}\right)^{n}$ which is isomorphic to $TH^2(M)$. 
This distinction is very important as it shows the difference between using $\bfQ$ and $\bfL_0^{-1}$. While $\bfQ$ is acting on elements of the torsion group, $\bfL_0^{-1}$ is acting on vectors in the $\bfL_0$ module $\left(\slfrac{\Z^{r}}{\mathbf{L}_{0}\Z^{r}}\right)^{n}$. When computing the partition function, this difference did not come into play. That is because we were summing over all elements of the torsion group $TH^2(M)$ and evaluating the bilinear forms on these elements only, so it did not matter which basis we were using. 
Moreover, the specific representative of the torsion class now also matters and it is part of the choices we have to make for this computation. On the expectation value side, this choice is only important for calculating the phase. In principle, we only need the class of $\boldsymbol{\ell}_\mathrm{right}$ for the sum. 

This might cause some confusion as to where each variable lies. For equation $\eqref{eq:lin_recip}$, on the right-hand-side sum, $\boldsymbol{\ell}_\mathrm{right}$ needs to only be defined $\mod (\bfL_0)^{\otimes n}$. However, on the phase factor, $\boldsymbol{\ell}'_\mathrm{middle}$ needs to be defined $\mod (\bfL_0 \otimes \bfK_0)$. Lastly, $\boldsymbol{\ell}'_\mathrm{left}$ is defined $\mod (\bfK_0)^{\otimes m}$. But of course, all these are just projections of a single $\boldsymbol{\ell}$ so it might seem paradoxical how it can be well defined in all these different modules. We will see the answer to that at the end of this section. 
%On the other hand, here, we also have a fixed element $\bht$ so the basis will matter. And so we need a way to obtain $\boldsymbol{\ell}_\mathrm{right}$ from $\bht$. In fact, that is not the right question. We will always have $\bht$ in some basis so we can express it, the actual question is, how do we convert $\bht$ from some basis of $\bfQ$ into the basis of $\bfL_0$. Since $\bfL_0$ can act as a linking form, there is an isomorphism $\phi$ between $\bfL_0$ and $\bfQ$ such that $\bfQ(a,b) = \bfL_0^{-1}(\phi a , \phi b)$. To get $\boldsymbol{\ell}_\mathrm{right}$ in such a case, we just need to do $\phi^{-1} \bht$.

Let's take a moment to understand the structure of $\boldsymbol{\ell}$ as a whole. In general, $\boldsymbol{\ell}$ contains some topologically trivial components but for this part of the computation we will not need them. A more accurate definition of $\boldsymbol{\ell}$ in the degenerate case would be $\boldsymbol{\ell}_i = \ell k(\bgamma, \mathcal{L}_i)|_{S^3}$. In $\mathrm{U}(1)$, $\ell$ has a part that comes from the torsion homology of the manifold and a part that comes from the free homology. In $\mathrm{U}(1)^n$, $\boldsymbol{\ell}$ has again the same components but $n$-fold (coming from $\bfK$ being $n\times n$), i.e. it is an $n$-dimensional vector where each component is a vector $\ell$ with a torsion and free part. During the computation, we see that the free part needs to be trivial $\mod\bfK$ and it does not contribute to the partition function so we only considered $\boldsymbol{\ell}_\mathrm{right} \simeq \boldsymbol{\tau}$, the torsion part. We see that this is a natural consequence of the partition function computation and it also appears naturally in the reciprocity formula. However, $\boldsymbol{\ell}_\mathrm{right}$ %_\mathrm{right} 
is also $n$-fold vector and some of its components might correspond to directions where $\bfK$ is degenerate. The structure of $\boldsymbol{\ell}$ similarly $\boldsymbol{\tau}$ can be shown in this diagram.
\[
\begin{tikzpicture}[baseline={(current bounding box.center)}]
\node (m) at (0,0) {$
\begin{array}{c|ccc|cc}
\boldsymbol{\ell}&&\bfL_{\mathrm{tors}}  && \bfL_{\mathrm{free}}  \\
     \hline 
     &&&&&\\
  \bfK_{\mathrm{tors}}&&\boldsymbol{\ell}_\mathrm{tors\; tors}  && \boldsymbol{\ell}_\mathrm{tors\; free} \\
  &&=\boldsymbol{\ell}_\mathrm{middle}&&\\
     \hline 
    \bfK_{\mathrm{free}} &&\boldsymbol{\ell}_\mathrm{free\; tors}  && \boldsymbol{\ell}_\mathrm{free\; free} \\
\end{array}
$};
\draw[red, thick, rounded corners] (-1.6,-1.2) rectangle (0.8,0.6)
node[below, midway, yshift=-25pt] {\footnotesize $=\boldsymbol{\ell}_{\mathrm{right}} \simeq \boldsymbol{\tau}$};
\draw[blue, thick, rounded corners] (-1.55,-0.65) rectangle (2.8,0.65)
node[below, midway, xshift=+80pt] {\footnotesize $=\boldsymbol{\ell}_{\mathrm{left}} $};
\end{tikzpicture}
\]%on From the Deloup-Turaev formula we have: 
\iffalse
We could also see this distinction with the following definitions:

\begin{align*}
     \boldsymbol{\ell}_\mathrm{middle} = \boldsymbol{\ell}_\mathrm{tors\; free}& =  \ell k(\bgamma_{\boldsymbol{t}}, \mathcal{L}_{tors})|_{S^3}\\
    \boldsymbol{\ell}_\mathrm{tors\; free}& =  \ell k(\bgamma_{\boldsymbol{t}}, \mathcal{L}_{tors})|_{S^3}.
\end{align*}
\fi
A priori there is no reason for the directions of $\boldsymbol{\ell}_{\mathrm{right}} \simeq \boldsymbol{\tau}$ corresponding to the degeneracy of $\bfK$ to be $0$. In the case where $\bfK$ is degenerate, we can always bring it to a form $\bfK_0 \oplus \mathbf{0}$ via a sequence of field redefinitions (the equivalent of the second Kirby moves for $\bfK$), where now $\mathbf{0}$ is some $0$-matrix equal to the dimension of the degenerate part of $\bfK$ (i.e its corank). Furthermore, let $\boldsymbol{\tau}_1  = \boldsymbol{\tau}|_{\bfK_\mathrm{tors}}\simeq \boldsymbol{\ell}_\mathrm{middle}$ and $\boldsymbol{\tau}_2 = \boldsymbol{\tau}|_{\bfK_\mathrm{free}}\simeq \boldsymbol{\ell}_\mathrm{free\; tors} $ to save space.  If $\bfK$ was an $n \times n$ matrix of rank $s$. The non-perturbative part of the expectation value will now become:
\begin{align*}
\sum_{\boldsymbol{\kappa}_{\mathbf{A}}\in\left(TH^{2}(M)\right)^{n}}
e^{-\pi i \boldsymbol{\kappa}_{\mathbf{A}} ^\top \left((\mathbf{K}_0 \oplus \mathbf{0}_{(n-s)}) \otimes \mathbf{Q}\right) \boldsymbol{\kappa}_{\mathbf{A}}-2\boldsymbol{\kappa}_{\mathbf{A}}^\top (\mathbf{Id}_n\otimes \mathbf{Q}) \boldsymbol{\tau}}  
\\
=\sum_{\boldsymbol{\kappa}_{\mathbf{A}}\in\left(TH^{2}(M)\right)^{n}}
\exp\left({-\pi i \boldsymbol{\kappa}_{\mathbf{A}} ^\top \left((\mathbf{K}_0 \oplus \mathbf{0}_{ (n-s)}) \otimes \mathbf{Q}\right) \boldsymbol{\kappa}_{\mathbf{A}}-2\boldsymbol{\kappa}_{\mathbf{A}}^\top ((\mathbf{Id}_{ s} \oplus \mathbf{Id}_{(n-s)})\otimes \mathbf{Q}) \boldsymbol{\tau}}\right)
\\
=\sum_{\boldsymbol{\kappa}_{\mathbf{A}}\in\left(TH^{2}(M)\right)^{s}}
\exp\left({-\pi i \boldsymbol{\kappa}_{\mathbf{A}} ^\top \left(\mathbf{K}_0  \otimes \mathbf{Q}\right) \boldsymbol{\kappa}_{\mathbf{A}}-2\boldsymbol{\kappa}_{\mathbf{A}}^\top (\mathbf{Id}_{s}\otimes \mathbf{Q}) \boldsymbol{\tau}_1}\right)
\\
\sum_{\boldsymbol{\kappa}_{\mathbf{A}}\in\left(TH^{2}(M)\right)^{n-s}}
\exp\left({-\pi i \boldsymbol{\kappa}_{\mathbf{A}} ^\top \left( \mathbf{0}_{(n-s)} \otimes \mathbf{Q}\right) \boldsymbol{\kappa}_{\mathbf{A}}-2\boldsymbol{\kappa}_{\mathbf{A}}^\top (\mathbf{Id}_{(n-s)}\otimes \mathbf{Q}) \boldsymbol{\tau}_2}\right).
\end{align*}
From the last result, the first sum is just the expectation value for the non-degenerate case. For the second sum, it is easy to see now that:
\[
\sum_{\boldsymbol{\kappa}_{\mathbf{A}}\in\left(TH^{2}(M)\right)^{n-s}}
\exp\left({-2\boldsymbol{\kappa}_{\mathbf{A}}^\top (\mathbf{Id}_{(n-s)}\otimes \mathbf{Q}) \boldsymbol{\tau}_2}\right) =\begin{cases}|TH^2(M) |^{n-s} & \text{if } \boldsymbol{\tau}_2 = 0  \, \text{(as a class)}%\mod (\bfL)^{(n-s)}
\\ 
0 &\text{otherwise}
\end{cases}
\]
With that in mind, we can now write the following about the non-perturbative part of our expectation value: 
\begin{align*}
    \sum_{\boldsymbol{\kappa}_{\mathbf{A}}\in\left(TH^{2}(M)\right)^{n}}&
e^{-\pi i \boldsymbol{\kappa}_{\mathbf{A}} ^\top \left((\mathbf{K}_0 \oplus \mathbf{0}_{(n-s)}) \otimes \mathbf{Q}\right) \boldsymbol{\kappa}_{\mathbf{A}}-2\boldsymbol{\kappa}_{\mathbf{A}}^\top (\mathbf{Id}_n\otimes \mathbf{Q}) \boldsymbol{\tau}}  
\\&= |TH^2(M) |^{n-s}\delta_{\boldsymbol{\tau}_2 } \sum_{\boldsymbol{\kappa}_{\mathbf{A}}\in\left(TH^{2}(M)\right)^{s}}
\exp\left({-\pi i \boldsymbol{\kappa}_{\mathbf{A}} ^\top \left(\mathbf{K}_0  \otimes \mathbf{Q}\right) \boldsymbol{\kappa}_{\mathbf{A}}-2\boldsymbol{\kappa}_{\mathbf{A}}^\top (\mathbf{Id}_{s}\otimes \mathbf{Q}) \boldsymbol{\tau}_1}\right)
\\&= \delta_{\boldsymbol{\tau}_2 } \sum_{\boldsymbol{\kappa}_{\mathbf{A}}\in\left(TH^{2}(M)\right)^{n}}
\exp\left({-\pi i \boldsymbol{\kappa}_{\mathbf{A}} ^\top \left((\mathbf{K}_0\oplus \mathbf{0}_{n-s})  \otimes \mathbf{Q}\right) \boldsymbol{\kappa}_{\mathbf{A}}-2\boldsymbol{\kappa}_{\mathbf{A}}^\top ((\mathbf{Id}_{s} \oplus \mathbf{0}_{n-s})\otimes \mathbf{Q}) \boldsymbol{\tau}}\right).
\end{align*}
In this form, this sum, matches the sum on the right hand side of $\ref{eq:lin_recip}$, thus we can use the reciprocity formula.

Furthermore, there is something extra we can say about $\boldsymbol{\ell}_\mathrm{free\; free}$. Earlier, we calculated that $\left( \mathbf{K}\mathbf{m}_{\mathbf{A}}+\mathbf{f} = 0\right)$ must have a solution for integer $\mathbf{m}_{\mathbf{A}}$. Now $\mathbf{f}$ is the equivalent of $\boldsymbol{\tau}$ for the free section, such that $\boldsymbol{\ell} \simeq \boldsymbol{\tau} \oplus \mathbf{f}$. For the degenerate directions of $\bfK$ this equation becomes $\mathbf{f}|_{\bfK_\mathrm{free}} = 0$. And since $\boldsymbol{\ell}_\mathrm{free\; free} = \mathbf{f}|_{\bfK_\mathrm{free}}$, then $\boldsymbol{\ell}_\mathrm{free\; free}  = 0$ exactly.  

Lastly, when $\bfK$ is degenerate, we will also have to deal with the cases where there are trivial observable loops on the directions that $\bfK$ is degenerate. For these directions, the integral:
\[
\int_{\left(\slfrac{\Omega^{1}(M)}{\Omega^{1}_{\mathrm{cl}}(M)}\right)^{s}}
\,\mathscr{D}\boldsymbol{\alpha}_{\perp}\,
e^{2\pi i\left(\int_{M}(\boldsymbol{\alpha}_{\perp})^\top \star\mathbf{C}\boldsymbol{\alpha}_{\perp}
+\int_M(\boldsymbol{\alpha}_{\perp})^\top \star \bh_{\boldsymbol{p}} \right)},
\]
ends up being
\[
\int_{\left(\slfrac{\Omega^{1}(M)}{\Omega^{1}_{\mathrm{cl}}(M)}\right)^{s}}
\,\mathscr{D}\boldsymbol{\alpha}_{\perp}\,
e^{2\pi i\left(\int_M(\boldsymbol{\alpha}_{\perp})^\top \star \boldsymbol{\omega} \right)} = 
\left(
\int_{\slfrac{\Omega^{1}(M)}{\Omega^{1}_{\mathrm{cl}}(M)}}
\,\mathscr{D}{\alpha}_{\perp}\,
e^{2\pi i\left(\int_M{\alpha}_{\perp} \wedge d\omega \right)} \right)^s.
\]
This expression is, of course, not mathematically well-defined, but we can perform a heuristic computation, just like in the partition function, to obtain a meaningful result.
The expression $2\pi i\int_M{\alpha}_{\perp} \wedge d\omega$ defines a linear functional on $\omega$; therefore, the whole expression can be considered an infinite-dimensional Fourier transform of the unit. To be consistent with the finite-dimensional cases, using the ideas of \cite{albeverio_infinite_2015,albeverio_theory_2006, thuillier_u1_2023}, we could extend the finite-dimensional Fourier transforms such that the whole expression can be:
\[\mbox{``} 
 \left(
\int_{\slfrac{\Omega^{1}(M)}{\Omega^{1}_{\mathrm{cl}}(M)}}
\,\mathscr{D}{\alpha}_{\perp}\,
e^{2\pi i\left(\int_M{\alpha}_{\perp} \wedge d\omega \right)} \right)^s =
\begin{cases}|\mathrm{vol}(\slfrac{\Omega^{1}(M)}{\Omega^{1}_{\mathrm{cl}}(M)})|^{s} & \text{if } \boldsymbol{\omega} = 0  \, \text{(as a class)}%\mod (\bfL)^{(n-s)}
\\ 
0 &\text{otherwise}
\end{cases}\mbox{''} 
\]
In other words, when $\boldsymbol{\omega}|_{\bfK_{\mathrm{free}}} = 0$, we will get an infinite factor that will get absorbed by the normalization and when $\boldsymbol{\omega}|_{\bfK_{\mathrm{free}}} \neq 0$, our partition function will be $0$. So overall, our partition function will get an extra $\delta_{\boldsymbol{\omega}|_{\bfK_{\mathrm{free}}}}$ term.

With everything in mind, we can now write $\ref{eq:final_obs}$ as
\begin{align}
\label{eq:final_final_obs}&
\nonumber 
\langle \langle W_M (\mathcal{L}, \boldsymbol{\gamma}|_{S^3}) \rangle \rangle_{\mathrm{CS}_{\mathbf{K}}}
=  |\det (\bfL_0)|^{n-s}\delta_{\boldsymbol{\omega}|_{\bfK_{\mathrm{free}}}}\delta_{\boldsymbol{\ell}_\mathrm{free\; free}}
\delta^{\mod \bfL_0}_{\boldsymbol{\ell}_\mathrm{free\; tors}}
\,
\delta^{\mod \bfK_0}_{\boldsymbol{\ell}_\mathrm{tors\; free}}
\,
e^{-\pi i (\bfK_0^{-1} \otimes \boldsymbol{\ell} \boldsymbol{k})(
   \boldsymbol{\gamma}_{\boldsymbol{0}},\boldsymbol{\gamma}_{\boldsymbol{0}})}
\\
&
   e^{ -\pi i \boldsymbol{\ell}_\mathrm{middle} ^\top (\bfK_0^{-1} \otimes \bfL_0^{-1})\boldsymbol{\ell}_\mathrm{middle}}.
\sum_{\boldsymbol{\kappa}_{\mathbf{A}}\in\left(\slfrac{\Z^{r}}{\mathbf{L}_0\Z^{r}}\right)^{s}}
e^{-\pi i \boldsymbol{\kappa}_{\mathbf{A}} ^\top \left(\mathbf{K}_0 \otimes \mathbf{L}_0^{-1}\right) \boldsymbol{\kappa}_{\mathbf{A}}-2\boldsymbol{\kappa}_{\mathbf{A}}^\top (\mathbf{Id}_s\otimes \mathbf{L}_0^{-1}) \boldsymbol{\ell}_\mathrm{middle} }
 \end{align}
%the partition function computed above will be different than the right hand side sum of the reciprocity formula. The difference is the partition function has the identity matrix tensored with the linking form while the reciprocity formula has an identity matrix directly summed with a $0$-matrix (equal to the dimension of the degenerate part, in other words the corank of $\bfK$). In these directions, in the partition function, the quadratic contribution will be $0$ (because of the degeneracy of $\bfK$) and we will have only the linear contribution. But as it happens when summing a linear term of that form, what we get is a Kronecker delta. In other words, the directions of $\boldsymbol{\tau}$ that correspond to the degenerate directions of $\bfK$ have to be $0\, \mathrm{mod} \bfL$. When this condition is satisfied, the rhs sums in equations \eqref{eq:lin_recip} and \eqref{eq:final_obs} are \textcolor{red}{equivalent}.
\\
\\
%%%%%%%%%%%%%%%%%%%%%%
\iffalse
\[
\begin{tikzpicture}[baseline={(current bounding box.center)}]
\node (m) at (0,0) {$
\begin{array}{c|ccc|cc}
\boldsymbol{\ell}&&\bfL_{\mathrm{tors}}  && \bfL_{\mathrm{free}}  \\
     \hline 
     &&&&&\\
  \bfK_{\mathrm{tors}}&&\boldsymbol{\ell}_\mathrm{tors\; tors}  && \boldsymbol{\ell}_\mathrm{tors\; free} \\
  &&=\boldsymbol{\ell}_\mathrm{middle}&&\\
     \hline 
    \bfK_{\mathrm{free}} &&\boldsymbol{\ell}_\mathrm{free\; tors}  && \boldsymbol{\ell}_\mathrm{free\; free} \\
\end{array}
$};
\draw[red, thick, rounded corners] (-1.6,-1.2) rectangle (0.8,0.6)
node[below, midway, yshift=-25pt] {\footnotesize $=\boldsymbol{\ell}_{\mathrm{right}} \simeq \bht$};
\draw[blue, thick, rounded corners] (-1.55,-0.65) rectangle (2.8,0.65)
node[below, midway, xshift=+80pt] {\footnotesize $=\boldsymbol{\ell}_{\mathrm{left}} $};
\end{tikzpicture}
\]%on From the Deloup-Turaev formula we have: 
\fi

%%%%%%%%%%%%%%%%%%%%%%%%%%%%%%%%5
By writing it like this, we can also notice, that in the last line, we can complete the square and get: 
%there is a term that corresponds to the perturbative part. If we combine this with the non-perturbative term, it completes the square and we would get:
\[
\sum_{\boldsymbol{\kappa}_{\mathbf{A}}\in\left(\slfrac{\Z^{r}}{\mathbf{L}_{0}\Z^{r}}\right)^{s}}
e^{-\pi i  (\boldsymbol{\kappa}_{\mathbf{A}}+ \bfK_0^{-1}\boldsymbol{\ell}_\mathrm{middle})^\top
\left(\mathbf{K}_0\otimes\mathbf{L}^{-1}_{0}\right)
(\boldsymbol{\kappa}_{\mathbf{A}} +\bfK_0^{-1}\boldsymbol{\ell}_\mathrm{middle})},
\]

with a small abuse of notation for $\bfK_0^{-1}\boldsymbol{\ell}_\mathrm{middle} $ to mean $ (\bfK_0^{-1}\otimes \mathbf{Id}_r)\boldsymbol{\ell}_\mathrm{middle}$. Now, with this form, we see that for the expectation value, $\boldsymbol{\ell}_\mathrm{middle}$ need only be defined $\mod (\bfK_0)^{\otimes r}$. This now shows how we can pass from a variable that is defined $\mod (\bfL_0)^{\otimes s}$ to one that is defined $\mod (\bfK_0)^{\otimes r}$. Finally, we can notice that before completing the square, the modulus of the sum on the last line, was dependent on $\boldsymbol{\ell}_\mathrm{middle} \mod (\bfL_0)^{\otimes s}$. After completing the square, the modulus depends on 
$\boldsymbol{\ell}_\mathrm{middle} \mod (\bfK_0)^{\otimes r}$, yet the only difference between them is a phase (which does not affect the modulus). This shows that the modulus in fact depends on both  $\boldsymbol{\ell}_\mathrm{middle} \mod (\bfL_0)^{\otimes s}$ and $\boldsymbol{\ell}_\mathrm{middle} \mod (\bfK_0)^{\otimes r}$. When $s=r=1$, $\mathbf{L}_0 = p$ and $\mathbf{K}_0 = q$, this just means that the modulus depends only on 
$\boldsymbol{\ell}_\mathrm{middle} \mod \gcd(p,q)$. On arbitrary dimensions, this would cause the modulus to depend on $\boldsymbol{\ell}_\mathrm{middle} \mod \mathbf{G}(\bfK_0^{\otimes r},\bfL_0^{\otimes s})$ with 
\[\mathbf{G}: \mathcal{M}_{d \times d}(\Z)\times \mathcal{M}_{d \times d}(\Z) \rightarrow \mathcal{M}_{d \times d}(\Z),\]
which would constitute a generalization of the $\gcd$ function from integers to integer matrices. 

Finally, we claim that this expectation value constitutes a manifold invariant. To show that, we just have to show invariance under Kirby moves on $\mathcal{L}$. As a reminder, two surgery links give the same manifold, if and only if they are related by a series of Kirby moves, ambient isotopy and reordering of link components. Since $\boldsymbol{\ell}$ is defined from the link, the Kirby moves will have an effect on it as well.
The first Kirby move, adds an unlinked, unknot with framing $\pm 1$ to the surgery link, effectively turning $\bfL$ into $\bfL \oplus \pm \mathbf{1}$. This component will be unlinked with the observable so it will add ``$0$-charged'' components to the appropriate positions on $\boldsymbol{\ell}$ (these positions are to be determined by the structure of each $\boldsymbol{\ell}$ but for an $n$-dimensional $\bfK$ matrix they will be added between the components $kn-1$ and $kn$ for integer $k$ if matrix element numbering starts from ``$1$'').
On the exponent of the expectation value, the new term will factorize such that:
\[
%\boldsymbol{\ell}_\mathrm{middle}
\left(\mathbf{K}_0\otimes\mathbf{L}^{-1}_{0}\right)
 \bfK_0^{-1}\boldsymbol{\ell}_\mathrm{middle} \rightarrow \left(\mathbf{K}_0\otimes\mathbf{L}^{-1}_{0}\right)
 (\bfK_0^{-1}\otimes \mathbf{Id}_r)\boldsymbol{\ell}_\mathrm{middle} \oplus \mathbf{K}_0\bfK_0^{-1} (0,\dots ,0)^\top .
\]
Therefore, the expectation value will factorize itself as:
\[
\langle \langle W_M (\mathcal{L}, \boldsymbol{\gamma}|_{S^3}) \rangle \rangle_{\mathrm{CS}_{\mathbf{K}}} \rightarrow \langle \langle W_M (\mathcal{L}, \boldsymbol{\gamma}|_{S^3}) \rangle \rangle_{\mathrm{CS}_{\mathbf{K}}}\cdot \sum_{\kappa_A \in (\Z / \Z)^s}e^{-\pi i\cdot0}= \langle \langle W_M (\mathcal{L}, \boldsymbol{\gamma}|_{S^3}) \rangle _{\mathrm{CS}_{\mathbf{K}}}.
\]
So the expectation value is invariant under the first Kirby move. There is one more thing to note regarding this Kirby move. When performing the inverse move, i.e., removing an unlinked (with the surgery link), unknotted component with framing $\pm 1$. If there are any observable loops linked to that, they do not disappear. Instead they become trivial loops of framing $\pm 1$ (since we had taken those to have framing $0$ pre-surgery). One might wonder where this framing comes from. The answer is that it arises from the surgery itself. The surgery on the $\pm 1$ component causes the space to twist around itself, without introducing any topology to the manifold but only causing a twist on the observable loop, changing its framing by one. 

For the second Kirby move, we will first explain the idea in the $\mathrm{U}(1)$ case and then show the general transformation. The second Kirby move from $i$ to $j$, ``adds'' the $i$-th surgery component to the $j$-th one. Essentially having $\mathcal{L}_j \rightarrow \mathcal{L}_j+\mathcal{L}_i$. This will cause the component of the observable link that previously coupled to $\mathcal{L}_i$ ($\ell_i$), to add its contribution to the component that couples to $\mathcal{L}_j$   
  ($\ell_j$). Essentially $\ell_j = \ell_j+\ell_i$. On the linking matrix, it can be represented by the transformation $\bfL \rightarrow P^\top \bfL P$, for a unimodular matrix $P$ \cite{moussard_realizing_2015,tagaris_u1u1_2023}. On $\ell$ the transformation will be $\ell \rightarrow P\ell$. In the $\mathrm{U}^n(1)$ case now, since $\boldsymbol{\ell}$ has $n$ number of observable links, the transformation will be $\boldsymbol{\ell} \rightarrow (\mathbf{Id}_n\otimes P) \boldsymbol{\ell}$. To calculate the effect on the expectation value, we will first note that as in the partition function, the transformation $\Z^r / \bfL_0 \Z^r  \rightarrow (P\Z^r)/\bfL_0 \Z^r$,  constitutes an automorphism of the group. And so, the sum must always be invariant under the transformation $\boldsymbol{\kappa}_{\mathbf{A}} \rightarrow (\mathbf{Id}_n\otimes P) \boldsymbol{\kappa}_{\mathbf{A}}$. So overall, we have that
  \begin{align*}
  \sum_{\boldsymbol{\kappa}_{\mathbf{A}}\in\left(\slfrac{\Z^{r}}{\mathbf{L}_{0}\Z^{r}}\right)^{s}}
e^{-\pi i  (\boldsymbol{\kappa}_{\mathbf{A}}+ \left(\mathbf{Id}_n\otimes P \right) \bfK_0^{-1}\boldsymbol{\ell}_\mathrm{middle})^\top
\left(\mathbf{K}_0\otimes\left(\left(P^{-1}\right)^\top\mathbf{L}^{-1}_{0}P^{-1}\right)\right)
(\boldsymbol{\kappa}_{\mathbf{A}} +\left(\mathbf{Id}_n\otimes P \right) \bfK_0^{-1}\boldsymbol{\ell}_\mathrm{middle})} = \\
\sum_{\boldsymbol{\kappa}_{\mathbf{A}}\in\left(\slfrac{\Z^{r}}{\mathbf{L}_{0}\Z^{r}}\right)^{s}}
e^{-\pi i  \left(\left(\mathbf{Id}_n\otimes P \right) \left(\boldsymbol{\kappa}_{\mathbf{A}} +\bfK_0^{-1}\boldsymbol{\ell}_\mathrm{middle}\right)\right)^\top
\left(\mathbf{K}_0\otimes\left(\left(P^{-1}\right)^\top\mathbf{L}^{-1}_{0}P^{-1}\right)\right)
\left(\left(\mathbf{Id}_n\otimes P \right) \left(\boldsymbol{\kappa}_{\mathbf{A}} +\bfK_0^{-1}\boldsymbol{\ell}_\mathrm{middle}\right)\right)} = \\
\sum_{\boldsymbol{\kappa}_{\mathbf{A}}\in\left(\slfrac{\Z^{r}}{\mathbf{L}_{0}\Z^{r}}\right)^{s}}
e^{-\pi i  \left(\boldsymbol{\kappa}_{\mathbf{A}} +\bfK_0^{-1}\boldsymbol{\ell}_\mathrm{middle}\right)^\top\left(\mathbf{Id}_n\otimes P \right)^\top 
\left(\mathbf{K}_0\otimes\left(\left(P^{-1}\right)^\top\mathbf{L}^{-1}_{0}P^{-1}\right)\right)
\left(\mathbf{Id}_n\otimes P \right) \left(\boldsymbol{\kappa}_{\mathbf{A}} +\bfK_0^{-1}\boldsymbol{\ell}_\mathrm{middle}\right)} =\\ 
\sum_{\boldsymbol{\kappa}_{\mathbf{A}}\in\left(\slfrac{\Z^{r}}{\mathbf{L}_{0}\Z^{r}}\right)^{s}}
e^{-\pi i  \left(\boldsymbol{\kappa}_{\mathbf{A}} +\bfK_0^{-1}\boldsymbol{\ell}_\mathrm{middle}\right)^\top
\left(\mathbf{K}_0\otimes\left(P^\top\left( P^{-1}\right)^\top\mathbf{L}^{-1}_{0}P^{-1} P\right)\right)
\left(\boldsymbol{\kappa}_{\mathbf{A}} +\bfK_0^{-1}\boldsymbol{\ell}_\mathrm{middle}\right)} = \\ 
\sum_{\boldsymbol{\kappa}_{\mathbf{A}}\in\left(\slfrac{\Z^{r}}{\mathbf{L}_{0}\Z^{r}}\right)^{s}}
e^{-\pi i  (\boldsymbol{\kappa}_{\mathbf{A}}+ \bfK_0^{-1}\boldsymbol{\ell}_\mathrm{middle})^\top
\left(\mathbf{K}_0\otimes\mathbf{L}^{-1}_{0}\right)
(\boldsymbol{\kappa}_{\mathbf{A}} +\bfK_0^{-1}\boldsymbol{\ell}_\mathrm{middle})}.
  \end{align*}
And so we see that the expectation value is invariant under the second Kirby move as well. Thus we have a manifold invariant.
\subsection{CS duality}
CS duality is the idea described in \cite{kim_u1n_2025}, that for specific descriptions of Chern-Simons (namely, the linking matrix being even which is always achievable), the partition function $\mathcal{Z}_{CS}(\mathbf{L},\mathbf{K})$ of a manifold described by linking matrix $\bfL$ and with coupling constant matrix $\bfK$ is related to the partition function $\mathcal{Z}_{CS}(\mathbf{K},\mathbf{L})$ by reciprocity formulas. 

We will now discuss if there is an indication of CS duality in the case of observables. First we have to understand what is our observable here as well. At the end of our computation, our observable was a link $\bgamma |_{S^3}$ living in $S^3$ that was interlinked with a surgery link $\mathcal{L}$ (described by the linking matrix $\bfL$), where we defined $\boldsymbol{\ell}_i = \ell k(\bgamma, \mathcal{L}_i)|_{S^3}$. We should not forget that in fact, $\boldsymbol{\ell}_i$ is itself a vector of dimension $n$, the dimension of the matrix $\bfK$. Note that the dual case is only defined if $\bfL$ is an even matrix, however, we can always transform $\bfL$ into an even matrix by a sequence of Kirby moves \cite{saveliev_lectures_1999, kim_u1n_2025}. We can write the above definition as $\boldsymbol{\ell}_{i}^j = \ell k(\bgamma^{j}, \mathcal{L}_i)|_{S^3}$ where the index $j \in \{0,\dots ,n-1\}$ refers to the part of $\bgamma$ belonging to the $j$-th copy of $\mathrm{U}(1)$. In the dual case, we would have a link $\mathcal{K}$ described by the linking matrix $\bfK$. The dual observable $\bgamma '$ will be subject to the following conditions 
\[
 \ell k((\bgamma')^{i}, \mathcal{K}_j)|_{S^3} = (\boldsymbol{\ell}')_j^i = \boldsymbol{\ell}_i^j = \ell k(\bgamma^{j}, \mathcal{L}_i)|_{S^3}.
\]
So where before $\bgamma|_{S^3}$ was interlinked with the components of $\mathcal{L}$, now $\bgamma'|_{S^3}$ is interlinked with the components of $\mathcal{K}$. We take the case where neither link contains trivial parts. With that in mind, the dual expectation value, would be:
\begin{align}
\label{eq:final_final_obs}&
\nonumber 
\langle \langle W_{M'} (\mathcal{K}, \boldsymbol{\gamma}'|_{S^3}) \rangle \rangle_{\mathrm{CS}_{\mathbf{L}}}
=  |\det (\bfK_0)|^{m-r}\delta_{\boldsymbol{\ell}_\mathrm{free\; free}}
\delta^{\mod \bfL_0}_{\boldsymbol{\ell}_\mathrm{free\; tors}}
\,
\delta^{\mod \bfK_0}_{\boldsymbol{\ell}_\mathrm{tors\; free}}
\,
%e^{-\pi i (\bfL_0^{-1} \otimes \boldsymbol{\ell} \boldsymbol{k})(  \boldsymbol{\gamma}'_{\boldsymbol{0}},\boldsymbol{\gamma}'_{\boldsymbol{0}})}
\\
&
   e^{ -\pi i {\boldsymbol{\ell}'}_\mathrm{middle} ^\top (\bfL_0^{-1} \otimes \bfK_0^{-1})\boldsymbol{\ell}'_\mathrm{middle}}.
\sum_{\boldsymbol{\kappa}_{\mathbf{A}}\in\left(\slfrac{\Z^{s}}{\mathbf{K}_0\Z^{s}}\right)^{r}}
e^{-\pi i \boldsymbol{\kappa}_{\mathbf{A}} ^\top \left(\mathbf{L}_0 \otimes \mathbf{K}_0^{-1}\right) \boldsymbol{\kappa}_{\mathbf{A}}-2\boldsymbol{\kappa}_{\mathbf{A}}^\top (\mathbf{Id}_r\otimes \mathbf{K}_0^{-1}) \boldsymbol{\ell}'_\mathrm{middle} }
 \end{align}
Then by using the reciprocity formulas we would have
\begin{align}
    \label{eq:obs_CS_duality}
    \nonumber e^{ \pi i {\boldsymbol{\ell}}_\mathrm{middle} ^\top (\bfK_0^{-1} \otimes \bfL_0^{-1}){\boldsymbol{\ell}}_\mathrm{middle}}{|\det (\bfK_0)|^{m-\frac{r}{2}}}
    &
\langle \langle W_M (\mathcal{L}, \boldsymbol{\gamma}|_{S^3}) \rangle \rangle_{\mathrm{CS}_{\mathbf{K}}}
  \\ =&
e^{-\frac{i\pi}{4}\left(\sigma\left(\mathbf{K}\right)\sigma\left(\mathbf{L}\right) \right)}|
\det (\bfL_0)|^{n-\frac{s}{2}}
%e^{-\pi i (\bfK_0^{-1} \otimes \boldsymbol{\ell} \boldsymbol{k})(\boldsymbol{\gamma}_{\boldsymbol{0}},\boldsymbol{\gamma}_{\boldsymbol{0}})}
\overline{
\langle \langle W_{M'} (\mathcal{K}, \boldsymbol{\gamma}'|_{S^3}) \rangle \rangle_{\mathrm{CS}_{\mathbf{L}}}}.
\end{align}
We see that again, just like in the partition function, the two expectation values are related.
\iffalse
\begin{align}
    \nonumber &
\langle \langle W_M (\mathcal{L}, \boldsymbol{\gamma}|_{S^3}) \rangle \rangle_{\mathrm{CS}_{\mathbf{C}}}
=  |\det (\bfL_0)|^{n-s}\delta_{\boldsymbol{\ell}_\mathrm{free\; free}}
\delta^{\mod \bfL_0}_{\boldsymbol{\ell}_\mathrm{free\; tors}}
\,
\delta^{\mod \bfK_0}_{\boldsymbol{\ell}_\mathrm{tors\; free}}
\,
e^{-\pi i (\bfK_0^{-1} \otimes \boldsymbol{\ell} \boldsymbol{k})(
   \boldsymbol{\gamma}_{\boldsymbol{0}},\boldsymbol{\gamma}_{\boldsymbol{0}})}
\\
&
   e^{ -\pi i \boldsymbol{\ell}_\mathrm{middle} ^\top (\bfK_0^{-1} \otimes \bfL_0^{-1})\boldsymbol{\ell}_\mathrm{middle}}
    \frac{1}{\abs{\determinant{\mathbf{K}_{0}}}^{\frac{r}{2}}}
   {\abs{\determinant{\mathbf{L}_{0}}}^{\frac{s}{2}}}
e^{-\frac{i\pi}{4}\left(\sigma\left(\mathbf{K}\right)\sigma\left(\mathbf{L}\right) \right)}\\
&\frac{1}{|\det (\bfK_0)|^{m-r}}\overline{
\langle \langle W_{M'} (\mathcal{K}, \boldsymbol{\gamma}'|_{S^3}) \rangle \rangle_{\mathrm{CS}_{\mathbf{L}}}}
\\
&
   ( e^{ +\pi i \boldsymbol{\ell}_\mathrm{middle} ^\top (\bfK_0^{-1} \otimes \bfL_0^{-1})\boldsymbol{\ell}_\mathrm{middle}})
\sum_{\boldsymbol{\kappa}_{\mathbf{A}}\in\left(\slfrac{\Z^{r}}{\mathbf{L}_0\Z^{r}}\right)^{s}}
e^{-\pi i \boldsymbol{\kappa}_{\mathbf{A}} ^\top \left(\mathbf{K}_0 \otimes \mathbf{L}_0^{-1}\right) \boldsymbol{\kappa}_{\mathbf{A}}-2\boldsymbol{\kappa}_{\mathbf{A}}^\top (\mathbf{Id}_s\otimes \mathbf{L}_0^{-1}) \boldsymbol{\ell}_\mathrm{middle} }
\end{align}
\fi
\newpage
\section{Examples}
In this section we will present a couple of examples that will hopefully help showcase the methodology.
\exple{
Let's see a comprehensive example on the expectation values of observables. We take our parameters to be the following matrices:
\[
\bfL = \begin{pmatrix}
    2 & 2 & -1\\
    2&2&-1 \\ 
    -1&-1&-2
\end{pmatrix},
\qquad
\bfK=
\begin{pmatrix}
    4 & 2 & 8\\
    2&4&4 \\ 
    8&4&16
\end{pmatrix}.
\]
We purposefully chose them to be non-invertible to showcase the various cases. It is not difficult to see that we can isolate an invertible part from a non-invertible in both cases. On $\bfL$ by a series of (second) Kirby moves, on $\bfK$ by a series of field redefinitions corresponding to sheer mappings like $A_1 \rightarrow A_1+A_2$ (note that algebraically, on the matrices the above moves and mappings are equivalent so the matrices can be treated in the exact same way). On $\bfK$ for example we could perform the mapping $A_2 \rightarrow A_2 - 2A_0$ (an algorithmic procedure on how to do this in the general case can be found in \cite{tagaris_u1u1_2023}). We can then turn both of these matrices into:
\[
\bfL \rightarrow \begin{pmatrix}
    -2 & 1 & 0 \\
    1 & 2 & 0 \\ 
    0 & 0 & 0 
\end{pmatrix}, \qquad \bfK \rightarrow
\begin{pmatrix}
    4 & 2 & 0\\ 
    2 & 4 & 0\\
    0 & 0 & 0
\end{pmatrix}.
\]
We can refer to the invertible (top-left) part of these matrices by $\bfL_0$ and $\bfK_0$. Note that this linking matrix could represent the manifold created by taking the connected sum of the lens space $L(5,2)$ with $S^2\times S^1$. It is time to pick our observable in $S^3$. Because of our earlier arguments we have the following: the trivial part of the observable can always be taken to be decoupled from the rest of the observable. The part of the observable coupling to the free sector can also taken to be decoupled from the rest of the observable (as it is always in the class of $0$). Finally, the torsion parts can be taken to be decoupled to each other, only coupling to one surgery link each and having framing $0$. The non-trivial part of the observable can thus be exhibited by $3$ (from the dimension of $\bfK$) independent $3$-component (from the dimension of $\bfL$) links coupled to the surgery link and so, can be fully described by a $9$-component vector (since $\bfK \otimes \bfL$ is $9\times 9$ dimensional). We will use the observable found in figure \ref{Fig:Example 1}. Note that the components of the observable that have charge $0 \mod \bfK$ can be ignored. Our vector will then be:
\[\boldsymbol{\ell} = \begin{pmatrix}
    1\\
    2\\ 
    0 \\
    -1\\
    4\\
    0\\
    0\\
    0\\
    0
\end{pmatrix}.\]
To showcase a bit better the projections of $\boldsymbol{\ell}$, we will present it in this $3 \times 3$ table.
\[
\begin{tikzpicture}[baseline={(current bounding box.center)}]
\node (m) at (0,0) {$
\begin{array}{c|ccc|cc}
\boldsymbol{\ell}&&\bfL_{\mathrm{tors}}  && \bfL_{\mathrm{free}}  \\
     \hline 
     &&&&&\\
  \bfK_{\mathrm{tors}}&&\begin{pmatrix}
      1 &2\\
      -1 & 4
  \end{pmatrix}  && \begin{pmatrix}
      0\\
      0
  \end{pmatrix}\\
     \hline 
    \bfK_{\mathrm{free}} &&\begin{pmatrix}
        0 & 0
    \end{pmatrix}  && \begin{pmatrix}
        0
    \end{pmatrix} \\
\end{array}
$};
\draw[red, thick, rounded corners] (-1.4,-1.2) rectangle (1,0.6)
node[below, midway, yshift=-25pt] {\footnotesize $=\boldsymbol{\ell}_{\mathrm{right}} \simeq \boldsymbol{\tau}$};
\draw[blue, thick, rounded corners] (-1.45,-0.57) rectangle (2.8,0.65)
node[below, midway, xshift=+80pt] {\footnotesize $=\boldsymbol{\ell}_{\mathrm{left}} $};
\end{tikzpicture}
\]
\begin{figure}
    \centering
    \begin{subfigure}[b]{0.4\textwidth}
         \centering         
         \begin{tikzpicture}
             \node[anchor=center,inner sep=0] at (0,0){\includegraphics[ width=1\linewidth]{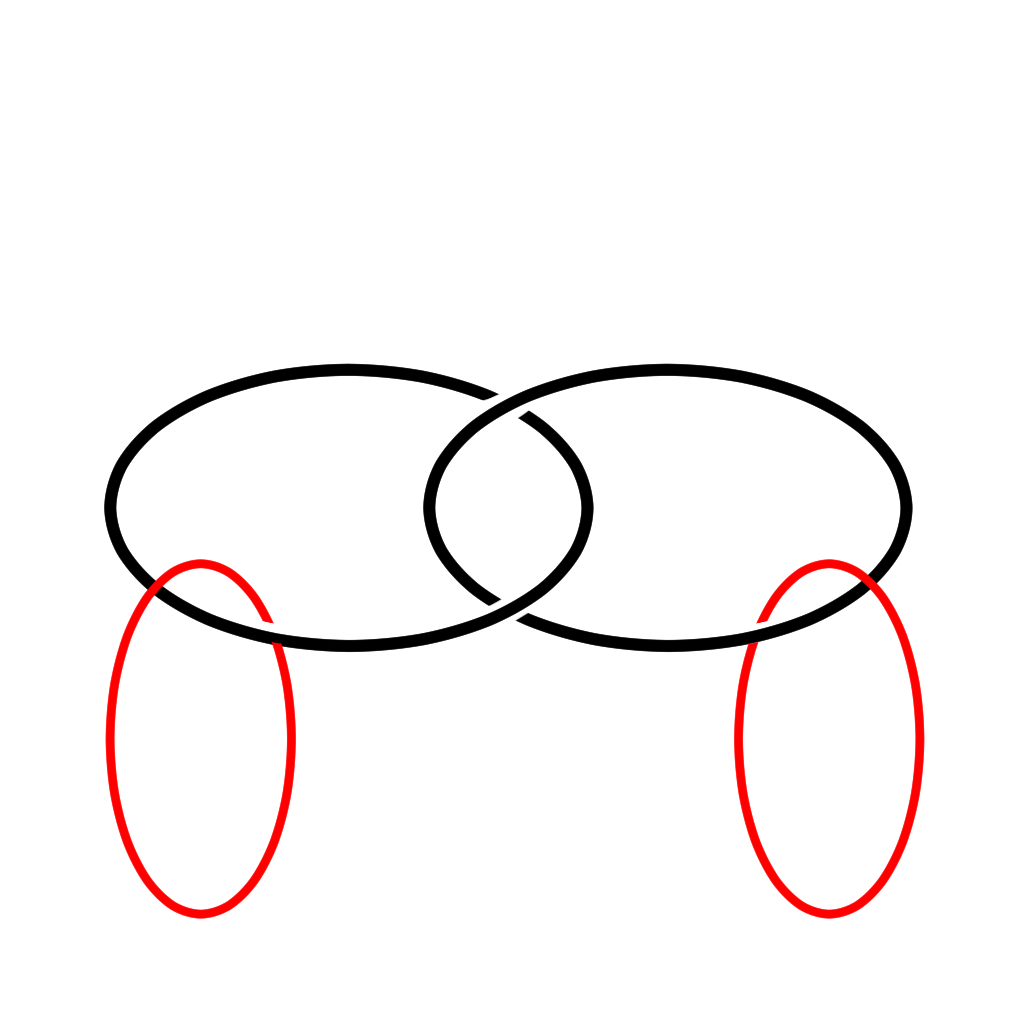}};
              \draw (-3,1.35)[color = black] node[right] {framing = $-2$};
              \draw (1,1.35)[color = black] node[right] {framing = $+2$};
              \draw (-3,-2.45)[color = red] node[right] {charge = $1$};
               \draw (1,-2.45)[color = red] node[right] {charge = $2$};
         \end{tikzpicture}
         \caption{The surgery link along with the non-perturbative observable in the \underline{first} copy of $\mathrm{U}(1)$.}
         \label{Fig:Ex1_first torsion}
     \end{subfigure}
     \hfill
     \begin{subfigure}[b]{0.4\textwidth}
         \centering         
         \begin{tikzpicture}
             \node[anchor=center,inner sep=0] at (0,0){\includegraphics[ width=1\linewidth]{Pictures/Ex1_tors.png}};
              \draw (-3,1.35)[color = black] node[right] {framing = $-2$};
              \draw (1,1.35)[color = black] node[right] {framing = $+2$};
              \draw (-3,-2.45)[color = red] node[right] {charge = $-1$};
               \draw (1,-2.45)[color = red] node[right] {charge = $4$};
         \end{tikzpicture}
         \caption{The surgery link along with the non-perturbative observable in the \underline{second} copy of $\mathrm{U}(1)$.}
         \label{Fig:Ex1_second torsion}
     \end{subfigure}
    
     \begin{subfigure}[b]{0.4\textwidth}
         \centering         
          \begin{tikzpicture}
             \node[anchor=center,inner sep=0] at (0,0){\includegraphics[ width=1\linewidth]{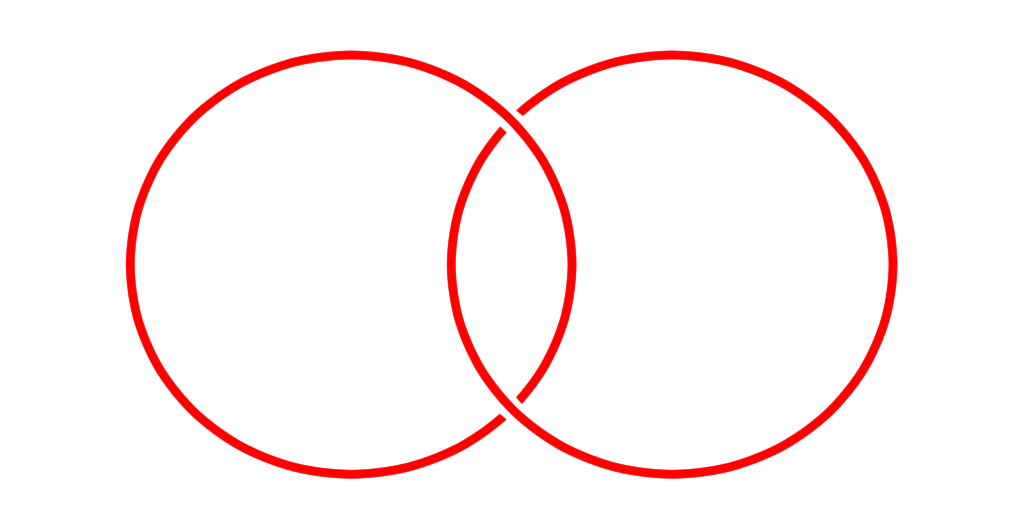}};
              \draw (0,2.55)[color = red] node[right] {};
              \draw (-3,1.55)[color = red] node[right] {framing = $0$};
              \draw (1,1.55)[color = red] node[right] {framing = $-4$};
              \draw (-3,-1.65)[color = red] node[right] {charge = $3$};
               \draw (1,-1.65)[color = red] node[right] {charge = $1$};
         \end{tikzpicture}
         \caption{The trivial part of the observable link, note that the two loops belong in different copies of $\mathrm{U}(1)$.}
         \label{Fig:trivial}
     \end{subfigure}
    
    \caption{Observable of example 1, we have split the figure into three parts. The first two each showcase the torsion component of the observable (in red) corresponding to a different copy of the gauge group, along with the surgery link (in black). The third part showcases the trivial component.}
    \label{Fig:Example 1}
\end{figure}
Furthermore, we will also add a trivial part to our observable, let's say a simple Hopf link whose components have charges $3$ and $1$, framings $0$ and $-4$, and live in the first and second copy of $\mathrm{U}(1)$ respectively. We can then start calculating $\eqref{eq:final_final_obs}$ for this case. To make the computation a little easier, we will go term by term.
\begin{align*}
    |\det(\bfL_0)|^{n-s} = |-5|^{3-2} &= 5,\\
    \delta_{\boldsymbol{\ell}_\mathrm{free\; free}}=
\delta^{\mod \bfL_0}_{\boldsymbol{\ell}_\mathrm{free\; tors}}
=
\delta^{\mod \bfK_0}_{\boldsymbol{\ell}_\mathrm{tors\; free}}&=1,\\
e^{\left(-\pi i \left(\bfK_0^{-1}\otimes \ell k \right)(\bgamma_{\boldsymbol{0}},\bgamma_{\boldsymbol{0}}) \right)} = e^{\left(-\pi i \left(\frac{1}{3}\cdot0\cdot 3^2+2\frac{-1}{6}\cdot1\cdot 3\cdot 1+\frac{1}{3}\cdot(-4)\cdot 1^2 \right) \right)} &= e^{\frac{\pi}{3}i}.
\end{align*}
Finally, we'll have the sum
\begin{align*}
\sum_{\boldsymbol{\kappa}_{\mathbf{A}}\in\left(\slfrac{\Z^{r}}{\mathbf{L}_{0}\Z^{r}}\right)^{s}}
e^{-\pi i ^\top (\boldsymbol{\kappa}_{\mathbf{A}}+ \bfK_0^{-1}\boldsymbol{\ell}_\mathrm{middle})
\left(\mathbf{K}_0\otimes\mathbf{L}^{-1}_{0}\right)
(\boldsymbol{\kappa}_{\mathbf{A}}+ \bfK_0^{-1}\boldsymbol{\ell}_\mathrm{middle})} = -5.
\end{align*}
So all in all,
\[
\langle \langle W_M (\mathcal{L}, \boldsymbol{\gamma}|_{S^3}) \rangle \rangle_{\mathrm{CS}_{\mathbf{K}}} = -25e^{\frac{\pi}{3}i}.
\]
}
\\
\exple{
In this example, we will showcase the CS duality. We could pick a simple lens space like $L(13,2)$
%$L(7,2)$ 
as our manifold, and a symmetrized coupling constant matrix such that:
\[
\bfL = \begin{pmatrix}
  %     -6 & 1\\
   %    1 & 2
    -2 & 1 & 0\\
   1 & 2  & 1\\
    0 & 1 & -2
\end{pmatrix}, \qquad \bfK = \begin{pmatrix}
8 & 3\\
3 & 4
\end{pmatrix}
\]
%with linking matrix $\bfL = \begin{pmatrix}
%    -3 & 1\\
%    1 & 2
%\end{pmatrix}$. And for our symmetrized coupling constant matrix we could pick $\bfK = \begin{pmatrix}
%    4 &3\\
%    3& 4
%\end{pmatrix}$. 
We also pick no free and trivial sectors for our observables and finally, $\boldsymbol{\ell} = (1,6,0,5,3,2)^\top$. So then we will have:
\[
\langle \langle W_M (\mathcal{L}, \boldsymbol{\gamma}|_{S^3}) \rangle \rangle_{\mathrm{CS}_{\mathbf{K}}} = 
\sum_{\boldsymbol{\kappa}_{\mathbf{A}}\in\left(\slfrac{\Z^{3}}{\mathbf{L}_{}\Z^{3}}\right)^{2}}
e^{-\pi i  (\boldsymbol{\kappa}_{\mathbf{A}}+ \bfK^{-1}\boldsymbol{\ell}_\mathrm{})^\top
\left(\mathbf{K}\otimes\mathbf{L}^{-1}_{}\right)
(\boldsymbol{\kappa}_{\mathbf{A}} +\bfK^{-1}\boldsymbol{\ell}_\mathrm{})}=12e^{-  \frac{16}{23}\pi i}.
\]
On the dual side, we would have
$\boldsymbol{\ell}' = (1,5,6,3,0,2)^\top$, so,
\[
\langle \langle W_{M'} (\mathcal{K}, \boldsymbol{\gamma}'|_{S^3}) \rangle \rangle_{\mathrm{CS}_{\mathbf{L}}} = 
\sum_{\boldsymbol{\kappa}_{\mathbf{A}}\in\left(\slfrac{\Z^{2}}{\mathbf{K}_{}\Z^{2}}\right)^{3}}
e^{-\pi i  (\boldsymbol{\kappa}_{\mathbf{A}}+ \bfK^{-1}\boldsymbol{\ell}'_\mathrm{})^\top
\left(\mathbf{L}\otimes\mathbf{K}^{-1}_{}\right)
(\boldsymbol{\kappa}_{\mathbf{A}} +\bfK^{-1}\boldsymbol{\ell}'_\mathrm{})}=(23)^{3/2}.
\]
And indeed we can confirm that:
\begin{align*}
     \nonumber e^{ \pi i {\boldsymbol{\ell}}_\mathrm{middle} ^\top (\bfK_0^{-1} \otimes \bfL_0^{-1}){\boldsymbol{\ell}}_\mathrm{middle}}{|\det (\bfK_0)|^{m-\frac{r}{2}}}
    &
\langle \langle W_M (\mathcal{L}, \boldsymbol{\gamma}|_{S^3}) \rangle \rangle_{\mathrm{CS}_{\mathbf{K}}}
  \\
=&e^{-\frac{37}{46}\pi i} \abs{23}^{3/2}\cdot12e^{-  \frac{16}{23}\pi i} \\
=& e^{  \frac{1}{2}\pi i} 12 \cdot  (23)^{3/2}
\\
=& e^{ - \frac{i\pi}{4}(2\cdot(-1))} \abs{12}^{2/2} \cdot  (23)^{3/2}
\\
=&
e^{-\frac{i\pi}{4}\left(\sigma\left(\mathbf{K}\right)\sigma\left(\mathbf{L}\right) \right)}|
\det (\bfL_0)|^{n-\frac{s}{2}}
%e^{-\pi i (\bfK_0^{-1} \otimes \boldsymbol{\ell} \boldsymbol{k})(\boldsymbol{\gamma}_{\boldsymbol{0}},\boldsymbol{\gamma}_{\boldsymbol{0}})}
\overline{
\langle \langle W_{M'} (\mathcal{K}, \boldsymbol{\gamma}'|_{S^3}) \rangle \rangle_{\mathrm{CS}_{\mathbf{L}}}}.
\end{align*}
}
\section{Conclusion}
The calculation of the expectation value of observables in $\mathrm{U}(1)$ Chern-Simons theories, naturally extends to the $\mathrm{U}(1)^n$ case, introducing new problems that needed to be addressed. Moreover, CS-duality exists for observables too and we show the relation between the dual expectation values. This article concludes our main study of $\mathrm{U}(1)^n$ CS on 3 dimensional closed oriented manifolds. In the future we wish to explore the case of manifolds with boundary and generalizing to higher dimensions. Specifically, the theory should work on $(4k+3)$-dimensional manifolds. However, the surgery approach implemented in this article does not extend to dimensions higher than $3$. Nevertheless, that will not be a problem since (as previously mentioned), the computation is ultimately done inside the manifold and can be performed irrespective of any Dehn surgery. 
\printbibliography

@article{moussard_realizing_2015,
    title = {Realizing isomorphisms between first homology groups of closed 3-manifolds by borromean surgeries},
    volume = {24},
    issn = {0218-2165},
    url = {https://www.worldscientific.com/doi/abs/10.1142/S0218216515500248},
    doi = {10.1142/S0218216515500248},
    number = {04},
    urldate = {2026-01-29},
    journal = {Journal of Knot Theory and Its Ramifications},
    author = {Moussard, D.},
    month = apr,
    year = {2015},
    pages = {1550024},
}

@article{albeverio_infinite_2015,
    title = {Infinite dimensional oscillatory integrals as projective systems of functionals},
    volume = {67},
    issn = {0025-5645, 1881-1167},
    url = {https://projecteuclid.org/journals/journal-of-the-mathematical-society-of-japan/volume-67/issue-4/Infinite-dimensional-oscillatory-integrals-as-projective-systems-of-functionals/10.2969/jmsj/06741295.full},
    doi = {10.2969/jmsj/06741295},
    number = {4},
    urldate = {2026-02-06},
    journal = {Journal of the Mathematical Society of Japan},
    author = {Albeverio, Sergio and Mazzucchi, Sonia},
    month = oct,
    year = {2015},
    pages = {1295--1316},
}

@inproceedings{albeverio_theory_2006,
    series = {Abel {Symposia}},
    title = {Theory and {Applications} of {Infinite} {Dimensional} {Oscillatory} {Integrals}},
    isbn = {978-3-540-70846-9},
    doi = {10.1007/978-3-540-70847-6_4},
    booktitle = {Stochastic {Analysis} and {Applications}, {Proceedings} of the {Abel} {Symposium} 2005 in
honor of {Prof}. {Kiyosi} {Ito},},
    publisher = {Springer, Berlin, Heidelberg},
    author = {Albeverio, Sergio and Mazzucchi, Sonia},
    year = {2006},
    pages = {75--92},
}

@book{saveliev_lectures_1999,
    title = {Lectures on the {Topology} of 3-manifolds: {An} {Introduction} to the {Casson} {Invariant}},
    isbn = {9783110162721},
    shorttitle = {Lectures on the {Topology} of 3-manifolds},
    language = {en},
    publisher = {Walter de Gruyter},
    author = {Saveliev, Nikolai},
    year = {1999},
    note = {Google-Books-ID: ErraOM8HYcIC},
}

@misc{thuillier_u1_2023,
    title = {The {U}(1) {BF} functional measure and the {Dirac} distribution on the space of quantum fields},
    url = {http://arxiv.org/abs/2306.04259},
    doi = {10.48550/arXiv.2306.04259},
    urldate = {2026-02-11},
    publisher = {arXiv},
    author = {Thuillier, Frank},
    month = jun,
    year = {2023},
    note = {arXiv:2306.04259 [math-ph]},
}

@phdthesis{mathieu_abelian_2018,
    type = {Theses},
    title = {Abelian {BF} theory},
    url = {https://theses.hal.science/tel-02132748},
    urldate = {2026-02-15},
    school = {Université Grenoble Alpes},
    author = {Mathieu, Philippe},
    month = jul,
    year = {2018},
}

@article{guadagnini_three-manifold_2013,
    title = {Three-manifold invariant from functional integration},
    volume = {54},
    issn = {0022-2488},
    url = {https://doi.org/10.1063/1.4818738},
    doi = {10.1063/1.4818738},
    number = {8},
    urldate = {2026-02-16},
    journal = {Journal of Mathematical Physics},
    author = {Guadagnini, Enore and Thuillier, Frank},
    month = aug,
    year = {2013},
    pages = {082302},
}

@article{deligne_theorie_1971,
    title = {Théorie de {Hodge} : {II}},
    volume = {40},
    issn = {1618-1913},
    shorttitle = {Théorie de {Hodge}},
    url = {https://www.numdam.org/item/?id=PMIHES_1971__40__5_0},
    language = {fr},
    urldate = {2026-02-16},
    journal = {Publications Mathématiques de l'IHÉS},
    author = {Deligne, Pierre},
    year = {1971},
    pages = {5--57},
}

@article{beilinson_higher_1985,
    title = {Higher regulators and values of {L}-functions},
    volume = {30},
    issn = {1573-8795},
    url = {https://doi.org/10.1007/BF02105861},
    doi = {10.1007/BF02105861},
    language = {en},
    number = {2},
    urldate = {2026-02-16},
    journal = {Journal of Soviet Mathematics},
    author = {Beilinson, Alexander A.},
    month = jul,
    year = {1985},
    pages = {2036--2070},
}

@misc{hossjer_extension_2022,
    title = {An extension of the \${\textbackslash}mathrm\{{U}\}{\textbackslash}!{\textbackslash}left(1{\textbackslash}right)\$ {BF} theory, {Turaev}-{Viro} invariant and {Drinfeld} center construction. {Part} {I}: {Quantum} fields, quantum currents and {Pontryagin} duality},
    shorttitle = {An extension of the \${\textbackslash}mathrm\{{U}\}{\textbackslash}!{\textbackslash}left(1{\textbackslash}right)\$ {BF} theory, {Turaev}-{Viro} invariant and {Drinfeld} center construction. {Part} {I}},
    url = {http://arxiv.org/abs/2212.12872},
    doi = {10.48550/arXiv.2212.12872},
    urldate = {2026-02-16},
    publisher = {arXiv},
    author = {Høssjer, Emil and Mathieu, Philippe and Thuillier, Frank},
    month = dec,
    year = {2022},
    note = {arXiv:2212.12872 [math-ph]},
}

@misc{hossjer_generalized_2023_a,
    title = {Generalized {Abelian} {Turaev}-{Viro} and $\mathrm{U}\!\left(1\right)$ {BF} {Theories}},
    url = {http://arxiv.org/abs/2302.09191},
    doi = {10.48550/arXiv.2302.09191},
    urldate = {2026-02-16},
    publisher = {arXiv},
    author = {Høssjer, Emil and Mathieu, Philippe and Thuillier, Frank},
    month = feb,
    year = {2023},
    note = {arXiv:2302.09191 [math-ph]},
}

@incollection{calaque_deligne-beilinson_2015,
    address = {Cham},
    title = {Deligne-{Beilinson} {Cohomology} in {U}(1) {Chern}-{Simons} {Theories}},
    isbn = {9783319099484 9783319099491},
    url = {https://link.springer.com/10.1007/978-3-319-09949-1_8},
    language = {en},
    urldate = {2026-02-17},
    booktitle = {Mathematical {Aspects} of {Quantum} {Field} {Theories}},
    publisher = {Springer International Publishing},
    author = {Thuillier, Frank},
    editor = {Calaque, Damien and Strobl, Thomas},
    year = {2015},
    doi = {10.1007/978-3-319-09949-1_8},
    pages = {233--271},
}

@book{brylinski_loop_1993,
    address = {Boston, MA},
    title = {Loop {Spaces}, {Characteristic} {Classes} and {Geometric} {Quantization}},
    copyright = {http://www.springer.com/tdm},
    isbn = {9780817647308 9780817647315},
    url = {http://link.springer.com/10.1007/978-0-8176-4731-5},
    language = {en},
    urldate = {2026-03-04},
    publisher = {Birkhäuser},
    author = {Brylinski, Jean-Luc},
    year = {1993},
    doi = {10.1007/978-0-8176-4731-5},
}

@incollection{esnault_deligne-beilinson_1988,
    address = {Boston},
    series = {Perspectives in {Mathematics}},
    title = {{DELIGNE}-{BEILINSON} {COHOMOLOGY}},
    volume = {4},
    url = {https://www.sciencedirect.com/science/chapter/edited-volume/pii/B9780125811200500094},
    language = {en-US},
    urldate = {2026-03-04},
    booktitle = {Beilinson's {Conjectures} on {Special} {Values} of {L}-{Functions}},
    publisher = {Academic Press},
    author = {Esnault, Hélène and Viehweg, Eckart},
    month = jan,
    year = {1988},
    doi = {10.1016/B978-0-12-581120-0.50009-4},
    pages = {43--91},
}

@article{guadagnini_path-integral_2014,
	title = {Path-integral invariants in abelian {Chern}–{Simons} theory},
	volume = {882},
	issn = {0550-3213},
	url = {https://www.sciencedirect.com/science/article/pii/S0550321314000832},
	doi = {10.1016/j.nuclphysb.2014.03.009},
	urldate = {2026-02-15},
	journal = {Nuclear Physics B},
	author = {Guadagnini, Enore and Thuillier, Frank},
	month = may,
	year = {2014},
	pages = {450--484},
}

@mastersthesis{tagaris_u1u1_2023,
	title = {U(1)×...×{U}(1) {Chern}-{Simons} theory},
	url = {https://www.research-collection.ethz.ch/entities/publication/36dadf9d-2b5c-4185-8bb1-c611048bc65c},
	doi = {10.3929/ethz-b-000623374},
	urldate = {2026-01-29},
	school = {ETH Zurich},
	author = {Tagaris, Michail},
	month = apr,
	year = {2023},
}

@incollection{mattes_invariants_1993,
	series = {Series on {Knots} and {Everything}},
	title = {On invariants of 3-manifolds derived from abelian groups},
	volume = {Volume 3},
	isbn = {9789810215446},
	url = {https://www.worldscientific.com/doi/10.1142/9789812796387_0018},
	number = {Volume 3},
	urldate = {2026-01-29},
	booktitle = {Quantum {Topology}},
	publisher = {WORLD SCIENTIFIC},
	author = {Mattes, Josef and Polyak, Michael and Reshetikhin, Nikolai},
	month = sep,
	year = {1993},
	pages = {324--338},
}

@misc{tagaris_reshetikhin-turaev_2025,
	title = {Reshetikhin-{Turaev} construction and {U}(1){\textasciicircum}n {Chern}-{Simons} partition function},
	url = {http://arxiv.org/abs/2507.08587},
	doi = {10.48550/arXiv.2507.08587},
	urldate = {2026-01-29},
	publisher = {arXiv},
	author = {Tagaris, Michail and Thuillier, Frank},
	month = jul,
	year = {2025},
	note = {arXiv:2507.08587 [math-ph]},
}

@article{kim_u1n_2025,
	title = {U(1){\textasciicircum}n {Chern}–{Simons} theory: {Partition} function, reciprocity formula and {Chern}–{Simons} duality},
	volume = {66},
	issn = {0022-2488},
	shorttitle = {U(1){\textasciicircum}n {Chern}–{Simons} theory},
	url = {https://doi.org/10.1063/5.0239253},
	doi = {10.1063/5.0239253},
	number = {4},
	urldate = {2026-01-29},
	journal = {Journal of Mathematical Physics},
	author = {Kim, Han-Miru and Mathieu, Philippe and Tagaris, Michail and Thuillier, Frank},
	month = apr,
	year = {2025},
	pages = {042301},
}

@article{deloup_reciprocity_2007,
	title = {On reciprocity},
	volume = {208},
	issn = {0022-4049},
	url = {https://www.sciencedirect.com/science/article/pii/S0022404905002963},
	doi = {10.1016/j.jpaa.2005.12.008},
	number = {1},
	urldate = {2026-01-29},
	journal = {Journal of Pure and Applied Algebra},
	author = {Deloup, Florian and Turaev, Vladimir},
	month = jan,
	year = {2007},
	pages = {153--158},
}
\section*{Appendix}
\appendix
\section{Equations of motion and zero modes} \label{Appendix}
\begin{figure}
\begin{center}
\includegraphics[scale = 1]{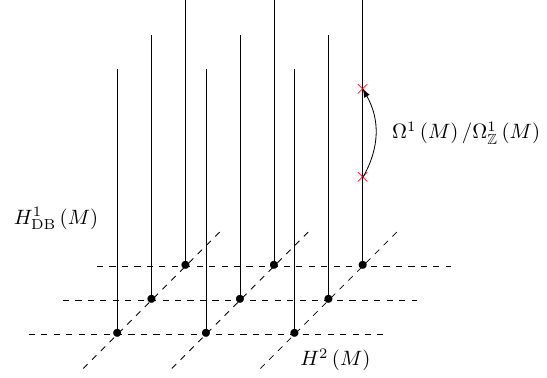}
\caption{Representation of $H^{1}_{\mathrm{DB}}(M)$ as a fibration with base space $H^2(M)$ and fibers $\Omega^1(M)/\Omega^1_\Z(M)$}
\label{Fig:space_fields}
\end{center}
\end{figure}

For the equations of motion, we will have to do a variation of the CS action as follows:
\[
    \delta S_{\mathrm{CS}_{\mathbf{C}}}[{A}] = \oint_M (\delta \,\,\mathbf{A} ^\top \star{\mathbf{C}} \mathbf{A} 
    + 
    \mathbf{A} \star{\mathbf{C}} \delta \mathbf{A}) = \oint_M (\delta \,\,\mathbf{A} ^\top \star (\mathbf{C} +  \mathbf{C}^\top)\mathbf{A}) = \oint_M \delta \,\,\mathbf{A} ^\top \star{\mathbf{K}} \mathbf{A}
\]
Immediately we get that an equation of motion would be along the directions corresponding to the kernel of $K$. 
\\
Now we want to examine what $\delta \mathbf{A}$ looks like. Reminder that the components $\mathbf{A}$ are classes $A$ belonging in $H^1_{DB}$. We can represent $H^1_{DB}$ as a fibration as seen in figure \ref{Fig:space_fields} \cite{hossjer_extension_2022}. 
\\
\\
Since the structure of $H^2(M)$ is discrete, it is easy to see that the variation of an $H^1_{DB}$ class along that direction would be $0$. So for $A \in H^1_{DB}$, $\delta A = (\delta \alpha , 0 , 0) $, where $\alpha$ is a global $1$-form. Then if $A = (A,\Lambda , n)$, $\delta A  \star A = \delta a \wedge dA$. 
Thus, in this case, the variation of the action is $0$ when $dA = 0$, that is, on the flat connections. 

So in total, the variation of $S_{\mathrm{CS}_{\mathbf{C}}}[{\mathbf{A}}]$ is $0$ when a direction is in the kernel of $K$ or it is a flat connection.
\\
Now for the zero modes. $\mathbf{Y}$ is a zero mode if the following is true for any field $\mathbf{A}$.
\[
S_{\mathrm{CS}_{\mathbf{C}}}[{\mathbf{A+\mathbf{Y}}}] = S_{\mathrm{CS}_{\mathbf{C}}}[{\mathbf{A\mathbf{}}}].
\]
In other words:
\begin{align*}
\oint_M (\mathbf{A}+\mathbf{Y})^\top \star{\mathbf{C}} (\mathbf{A}+\mathbf{Y})  &= \oint_M \mathbf{A} ^\top \star{\mathbf{C}} \mathbf{A}\\
\oint_M \left(\mathbf{A} ^\top \star{\mathbf{C}} \mathbf{A}
+\mathbf{A}^\top \star{\mathbf{C}} \mathbf{Y} + \mathbf{Y}^\top \star{\mathbf{C}} \mathbf{A}  + \mathbf{Y}^\top \star{\mathbf{C}} \mathbf{Y}  \right) &= \oint_M \mathbf{A} ^\top \star{\mathbf{C}} \mathbf{A}\\
\oint_M \left(\mathbf{A}^\top \star{\mathbf{K}} \mathbf{Y} +\mathbf{Y}^\top \star{\mathbf{C}} \mathbf{Y}  \right) &= 0.
\end{align*}
We start first with $\mathbf{A}^\top \star{\mathbf{K}} \mathbf{Y} $. The first part in every component of that term will be something of the form $A\wedge dY$. For general $A$, this can only be $0$ if $dY = 0$. So $Y$, by the 2nd decomposition, can be expressed as $Y = y_0 + k_Y$ with $y_0 \in \Omega^1_{cl} / \Omega^1_\Z$ and $k_Y \in TH^2$. We need for them, independently $\mathbf{A}^\top \star \mathbf{K} \boldsymbol{y}_0 = 0$ and $\mathbf{A} ^\top \star \mathbf{K} \boldsymbol{k}_\mathbf{Y} =0$ where now $\boldsymbol{y}_0 $ and $\boldsymbol{k}_\mathbf{Y}$ are the tuples containing the $y_0$ and $k_Y$ components respectively. This means that $\mathbf{K}\boldsymbol{y}_0  \sim0$ and $\mathbf{K} \boldsymbol{k}_\mathbf{Y} \sim 0$ in terms of classes. \\\\
For the $\mathbf{Y}^\top \star{\mathbf{C}} \mathbf{Y} $ term we use the same decomposition again. We note that using good origins for our fibers \cite{hossjer_generalized_2023_a}, $y_0 \star y_0=0$ and $y_0 \star k_Y = 0$. Therefore, the only term remaining would be $k_Y \star k_Y$. This is the usual linking form and can be rewritten as $(\bfQ \otimes \mathbf{C})(\boldsymbol{k}_\mathbf{Y})$. So we get another condition: $(\bfQ \otimes \mathbf{C})(\boldsymbol{k}_\mathbf{Y}) = 0$. It is important to note that this condition is not implied by the earlier condition $\mathbf{K}\boldsymbol{k}_\mathbf{Y} \sim 0$ since from that we would get that $0 = \mathbf{Y}^\top \star{\mathbf{K}} \mathbf{Y}  = \mathbf{Y}^\top \star{(\mathbf{C}+\mathbf{C}^\top)} \mathbf{Y} = 2\mathbf{Y}^\top \star{\mathbf{C}} \mathbf{Y}$.
\\\\
For $y_0$ this means that $y_0 = \mathbf{K}^{-1}R$ where $R$ is a vector whose components are $(\rho , 0 , 0)$ where $\rho \in \Omega^1_\Z$. 
\\
For $\boldsymbol{k}_\mathbf{Y}$ the idea is similar but it is more restrictive. We have to find specific torsion classes $\boldsymbol{k}_\mathbf{Y}$ such that $\mathbf{K} \boldsymbol{k}_\mathbf{Y} =0$. Unlike for the $y_0$, here we cannot take any torsion class equivalent to $0$ and multiply that by $\mathbf{K}^{-1}$ as it will not in general produce a torsion class. Then out of those classes, pick the ones that fulfill $(\bfQ \otimes \mathbf{C})(\boldsymbol{k}_\mathbf{Y}) = 0$. For this reason, this zero mode is more of a coincidence and it is important to distinguish it from the $y_0$ zero mode. An example would be in the simple $U(1)$ case with $\mathbf{C} = 1$, $TH^2 = \Z_4$ and $\bfQ = 1/4$ where we have $g_4$ a  $\Z_4$ generator and $\boldsymbol{k}_\mathbf{Y} = 2g_4$. In this case, $\mathbf{K} \boldsymbol{k}_\mathbf{Y} = 4g_4 \sim0$  and $(\bfQ\otimes \mathbf{C}) (\boldsymbol{k}_\mathbf{Y}) = 2\frac{1}{4}2 =1\sim 0$.

\end{document}